%% file: policy-generation.tex
\documentclass[lettersize,journal]{IEEEtran}
\usepackage{amsmath,amsfonts}
\usepackage{algorithmic}
\usepackage{algorithm}
\usepackage{array}
\usepackage[caption=false,font=normalsize,labelfont=sf,textfont=sf]{subfig}
\usepackage{textcomp}
\usepackage{stfloats}
\usepackage{url}
\usepackage{verbatim}
\usepackage{graphicx}
\usepackage{cite}
\usepackage{tikz}
\usepackage{pifont}
\usepackage{fontawesome}
\usepackage{xcolor}
\usepackage{booktabs} % For formal tables
\usepackage{color}
\usepackage{soul}
\usepackage{multirow}
\usepackage{multicol}
\usepackage{pdflscape}
\usepackage{mathrsfs}
\usepackage{amsbsy}
\usepackage{tcolorbox}
\usepackage{balance}
\usepackage{microtype}
\usepackage{longtable}
\usepackage{indentfirst}
\usepackage{pgfplots}
\usepackage{enumitem}
\usepackage{eurosym}
\usepackage[numbers]{natbib}
\usepackage[framemethod=TikZ]{mdframed}
\usepackage{xurl}
\mdfsetup{skipabove=3pt,skipbelow=1pt}

\newcommand{\newtext}[1]{\textcolor{black}{#1}}

\begin{document}
	
\title{Interactive GDPR-Compliant Privacy Policy Generation for Software Applications}

\author{Pattaraporn Sangaroonsilp, Hoa Khanh Dam, Omar Haggag, John Grundy}

%\thanks{P. Sangaroonsilp, H. K. Dam and A. Ghose are with the School of Computing and Information Technology, Faculty of Engineering and Information Sciences, University of Wollongong, Australia. \hfil\break Email: ps642@uowmail.edu.au and \{hoa, aditya\}@uow.edu.au

\markboth{Journal of \LaTeX\ Class Files,~Vol.~14, No.~8, August~2021}%
{Shell \MakeLowercase{\textit{et al.}}: A Sample Article Using IEEEtran.cls for IEEE Journals}

\IEEEpubid{0000--0000/00\$00.00~\copyright~2021 IEEE}

\maketitle
	
\begin{abstract}
\newtext{Software applications are designed to assist users in conducting a wide range of tasks or interactions. They have become prevalent and play an integral part in people's lives in this digital era. To use those software applications, users are sometimes requested to provide their personal information. As privacy has become a significant concern and many data protection regulations exist worldwide, software applications must provide users with a privacy policy detailing how their personal information is collected and processed. We propose an approach that generates a comprehensive and compliant privacy policy with respect to the General Data Protection Regulation (GDPR) for diverse software applications. To support this, we first built a library of privacy clauses based on existing privacy policy analysis. We then developed an interactive rule-based system that prompts software developers with a series of questions and uses their answers to generate a customised privacy policy for a given software application. We evaluated privacy policies generated by our approach in terms of readability, completeness and coverage and compared them to privacy policies generated by three existing privacy policy generators and a Generative AI-based tool. Our evaluation results show that the privacy policy generated by our approach is the most complete and comprehensive.}
\end{abstract}


\input{sections/1-intro}
\input{sections/2-related-work}
\input{sections/3-approach-overview}
\input{sections/4-clauses-library}
\input{sections/4-clauses-refinement}
\input{sections/5-rule-basedQA}
\input{sections/6-evaluation}
\input{sections/7-threats}
\input{sections/8-conclusion}

\balance
\bibliographystyle{IEEEtranN}
\bibliography{Privacy-policy,ICSE2020_kookai_ref,privacy-requirements}

\end{document}

%% file: sections/1-intro.tex
\section{Introduction} \label{sec:intro}

Software applications are developed to perform tasks or facilitate activities for users. With the pervasive integration of technology into daily life, many of these applications collect and process personal data and sensitive information. As privacy concerns have escalated and many data protection regulations have been enacted worldwide (e.g., the European Union's General Data Protection Regulation (GDPR) \cite{OfficeJournaloftheEuropeanUnion;2016} and California Consumer Privacy Act (CCPA) \cite{StateofCaliforniaDepartmentofJustice2018}), software developers must ensure that they inform users about their individual rights and how personal information is collected and processed within the applications. This information is communicated through a privacy policy.

A privacy policy is a document that explains to general users how an organisation processes personal data (e.g., collection, use, storage, dissemination and transfer) and must comply with specific data protection and/or privacy regulations \cite{Amaral2021}. It also describes how a software application satisfies privacy-related requirements. Privacy policies provide transparency for users so that they can make decisions before providing their personal data to software applications. \emph{Due to the varying scope and levels of protection provided by data protection and privacy legislation in different jurisdictions, organisations face challenges in developing a privacy policy that aligns with the functionalities and features of their software applications while also complying with the necessary regulations.}

\newtext{Many challenges concerning privacy policies have been identified and investigated, including the difficulties developers encounter when developing privacy policies and incomplete or non-compliant privacy policies. \citeauthor{Tahaei2023} discussed the challenges developers face in integrating privacy considerations into the design and development process, resulting in unclear and ineffective privacy policies \cite{Tahaei2023}.} Previous studies have proposed various approaches to assess the completeness and compliance of the privacy policies against specific data protection and privacy legislations \cite{Guntamukkala2016, Tesfay2018, Torre, Amaral2021, Muller2019, Anthonysamy2014}. However, they assumed that privacy policies were already available and did not provide any support to create a customised privacy policy. Some web-based tools provide free and paid services for privacy policy generation (e.g. \cite{Termly2022, PrivacyPolicies2022, TermsFeed2023}). However, privacy policies generated from those tools are not tailored to specific software applications and their unique functionalities. \newtext{Hence, there is a significant gap in providing adequate support for software developers and privacy policymakers to generate complete and comprehensive privacy policies that align with the functionalities and features of their software applications while also ensuring compliance with data protection regulations.}

In this paper, we propose an approach, PPGen, designed to generate privacy policies that comply with GDPR for any software application. GDPR is selected due to its wide recognition as one of the world's most well-established, widely adopted and overarching data protection regulations in the world. Although the current work is scoped to GDPR, our approach can be readily extended to other privacy regulations. Our approach consists of two processes: privacy clause development and privacy policy generation. \newtext{In the first phase, we identified and extracted common privacy policy clauses that are required for privacy policies for software applications. We then identified the privacy clauses that comply with GDPR based on metadata types (see Section \ref{sec:clause-development} for more details). Finally, we extracted and refined those clauses and constructed a library of privacy clauses. In the second phase, we created a prototype tool for the automated generation of privacy policies. To generate a privacy policy, the system presents a set of questions to a software developer. The questions are asked based on the answers inputted by the developer in the previous questions. The PPGen system then generates a customised privacy policy for that software application. \IEEEpubidadjcol We employed three evaluation criteria (readability, completeness and coverage) to evaluate the privacy policies generated by the PPGen system and compared its performance against three online policy generators and a Generative AI-based tool.}

This work makes the following key contributions:

\begin{enumerate}
    \item \newtext{a framework (PPGen) for constructing a comprehensive library of privacy clauses and generating privacy policies;}
    \item \newtext{a prototype tool for generating GDPR-compliant privacy policies for software applications; and}
    \item \newtext{an evaluation of our approach showing that it generates high-quality GDPR-compliant privacy policies for both existing and new software applications and outperforms existing privacy policy generation and Generative AI-based tools.}
\end{enumerate}

A full replication package containing all the artifacts and results generated by our study is available at \cite{reppkg-privacygen}. The remainder of the paper is structured as follows. Section \ref{sec:related-work} discusses related work before an overview of our approach is presented in Section \ref{sec:approach-overview}. Section \ref{sec:clause-development} discusses how a library of privacy clauses was developed, while Section \ref{sec:rule-based} explains the steps used to develop a system for generating privacy policies. The evaluation criteria and results are reported in Section \ref{sec:evaluation}. Section \ref{sec:threats} discusses the threats to validity of our work. We conclude our work in Section \ref{sec:conclusion}.

%% file: sections/2-related-work.tex
\section{Related Work} \label{sec:related-work}

Generating a privacy policy is a time-consuming and error-prone task for application developers \cite{Sun2020}. It requires a significant effort from authors and stakeholders to ensure that the policy is correct and aligns with the functionalities provided by the software applications. In addition, it is required to provide policies that comply with relevant data protection regulations and privacy laws (e.g., GDPR) if they process personal data. However, there is limited work supporting the generation of customised privacy policies for software applications. 
\citeauthor{Yu2017} \cite{Yu2017, Yu2015a, Yu2021} developed a system to facilitate the privacy policy generation for Android applications. Their system employed static code analysis to characterise the apps' behaviours related to the users' personal information. It then uses natural language processing (NLP) techniques to automatically generate privacy policies based on those behaviours. \citeauthor{Miao2014} \cite{Miao2014} proposed a solution to automatically generate privacy descriptions for mobile applications by analysing source code in a mobile application building platform called App Inventor. 
\citeauthor{Rowan2014} \cite{Rowan2014} implemented a plugin tool that enables application developers to create privacy policies while developing their applications in the Eclipse integrated development environment (IDE). Their tool provides a set of questions for the developers to answer instead of analysing the apps' source code. However, none of the above studies has investigated the regulatory compliance aspect of privacy policies.

Online privacy policy generators provide free and paid services for generating privacy policies for web and mobile applications (e.g., \cite{Termly2022, PrivacyPolicies2022, Shop}). Those tools ask a set of common questions related to the collection and use of personal data (e.g., ``Who are the users of the application?'', ``What kind of personal data do you collect from the users?'' and ``For which purposes will the personal data be used?''). A user needs to provide answers to those questions to generate a privacy policy. Those tools also provide the options for the users to select the specific privacy regulation(s) they would like their applications to comply with. However, the privacy policy generated from those tools contains generic statements that are not customised for a certain software application and its unique functionalities.

\citeauthor{Sun2020} \cite{Sun2020} assessed the quality of ten online Automated Privacy Policy Generators (APPGs). Their study summarised three interesting findings. Firstly, some APPGs rely heavily on templates to generate privacy policies, which do not support the different functionalities of different applications. Secondly, some APPGs generated statements that are over-scoped from the ground truth inputted by the user. Finally, the questionnaires developed by some APPGs are too complex for the user to answer. Some of the questions did not allow the user to answer in a more flexible way. %We aim to mitigate these limitations in our work.

Recent work has proposed AI-assisted approaches for checking the completeness of privacy policies. \citeauthor{Torre} \cite{Torre} built a conceptual model of privacy-policy metadata types based on the privacy-related provisions of GDPR. The metadata types are the components that need to be presented in the privacy policies. Those metadata types were then used to create a set of criteria to check the completeness of privacy policies against GDPR. In addition, this work also adopted NLP and supervised machine learning (ML) to develop an automated solution for completeness checking. The authors evaluated their approach on real privacy policies from the fund industry. \citeauthor{Amaral2021} \cite{Amaral2021} extended the work proposed in \cite{Torre} by providing more concrete and detailed examples, applying the AI-based approach to identify all the 56 metadata types in a given privacy policy (from 20 metadata types in the previous work) and adding more privacy policies in the evaluation process. %We use the metadata types proposed in \cite{Amaral2021} to extract relevant privacy clauses from the privacy policies of OLAs in this paper.

\citeauthor{Guntamukkala2016} \cite{Guntamukkala2016} proposed an automated approach using both ML and text mining techniques to evaluate the completeness of online privacy policies. This work considered the completeness based on the presence of eight main categories suggested by the United States Federal Trade Commission (FTC) Fair Information Practices Principles (FIPP) and Organisation for Economic Co-operation and Development (OECD) guidelines. However, their work did not generate privacy policies.

%% file: sections/3-approach-overview.tex
\section{PPGen Approach} \label{sec:approach-overview}

\begin{figure}[ht]
    \centering
    \includegraphics[width=1.0\linewidth]{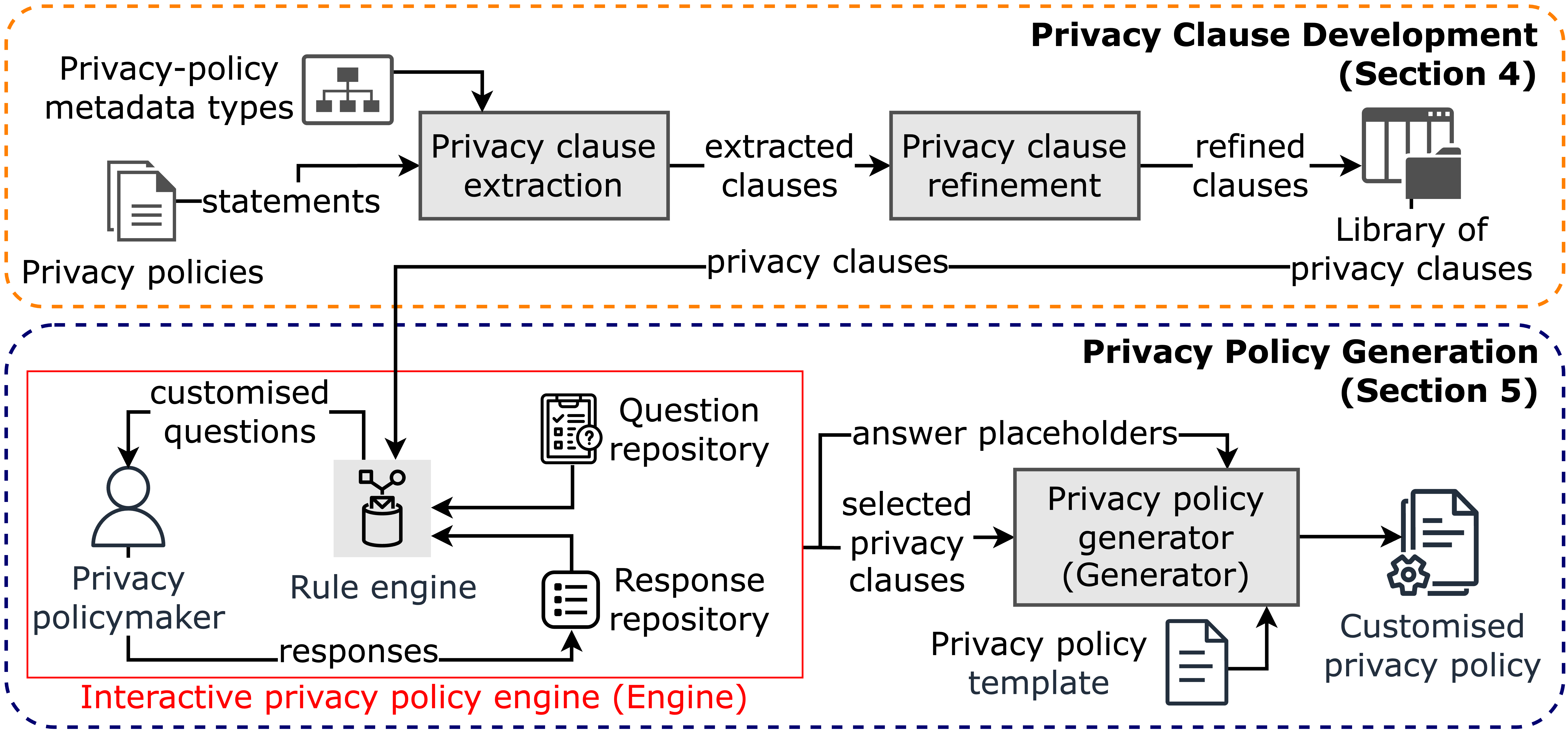}
    \caption{PPGen framework}
    \label{fig:PPGen-framework}
\end{figure}

Our PPGen framework consists of two processes: \emph{privacy clause development} and \emph{privacy policy generation} (see Figure \ref{fig:PPGen-framework}). There are two modules in the privacy clause development: privacy clause extraction and refinement. The former extracts statements from existing privacy policies and then uses them to form privacy clauses and classify them based on privacy-policy metadata types proposed in \cite{Amaral2021}. Those privacy clauses are fed into the privacy clause refinement module to remove inconsistencies and duplicates. After the refinement, the refined clauses are stored in \emph{privacy clause library}. 

The privacy policy generation process consists of two modules: \emph{interactive privacy policy engine (Engine)} and \emph{privacy policy generator (Generator)}. The Engine module has five components: question repository, response repository, privacy clause library, rule engine and privacy policymaker. The question repository contains a list of predefined questions, while the response repository has a list of predefined responses and answer placeholders, which will be later presented to a privacy policymaker. The rule engine determines the association between questions, responses and privacy clauses and controls the question flow. It also determines a series of customised questions to ask the privacy policymaker based on their responses to the previous questions. The responses provided by the privacy policymaker are stored in the response repository and are fed into the rule engine. The Engine module then transmits a set of selected privacy clauses and answer placeholders to the Generator module. The Generator module generates a customised privacy policy by placing the selected privacy clauses and answers placeholders based on a privacy policy template. Our framework is generic and applicable to generate a customised privacy policy for any software application, as every component is adaptable. Section \ref{sec:clause-development} and Section \ref{sec:rule-based} elaborate on the privacy clause development and privacy policy generation processes in detail through a case study of online learning applications (OLAs).

%% file: sections/4-clauses-library.tex
\section{Privacy clause development} \label{sec:clause-development}

This section explains the process of constructing a library of privacy clauses, which involves two key steps: privacy clause extraction and privacy clause refinement. %We showcase this process using the selected OLAs. A library of 125 privacy clauses categorised into 56 categories was obtained from the privacy policies of ten existing OLAs.

%\newtext{In this study, we use online learning platforms (OLAs) as a case study.} Online learning has emerged as an alternative to mitigate the limitations and difficulties of traditional learning (e.g., location issues and time differences) and has become increasingly prevalent on account of the COVID-19 pandemic \cite{Lockee2021, Yan2021, Chandrasiri2022}. OLAs such as Moodle \cite{Moodlea}, Coursera \cite{Coursera} and edX \cite{edX2021} have become widely used, not only for general studies outside of classrooms but also in universities and secondary education systems. Before using OLAs, users must provide a variety of personal data (e.g., name, address, email address and credit card information). This poses a threat to user privacy if the users are not well informed of how their personal data is collected and processed. 
%Thus, in this work, we analysed the privacy policies of ten popular OLAs, including Moodle \cite{Moodlea}, Coursera \cite{Coursera}, edX \cite{edX2021}, FutureLearn \cite{FutureLearn2022}, Udemy \cite{Udemy2021}, Brilliant \cite{Brilliant2020}, MasterClass \cite{MasterClass2020}, Mindvalley \cite{Mindvalley2021}, SkillShare \cite{Skillshare2019} and Udacity \cite{Udacity2022}. 

\subsection{Privacy clause extraction} \label{sec:clause-extraction}

\subsubsection{Metadata} 
\begin{figure}
    \centering
    \includegraphics[width=0.9\linewidth]{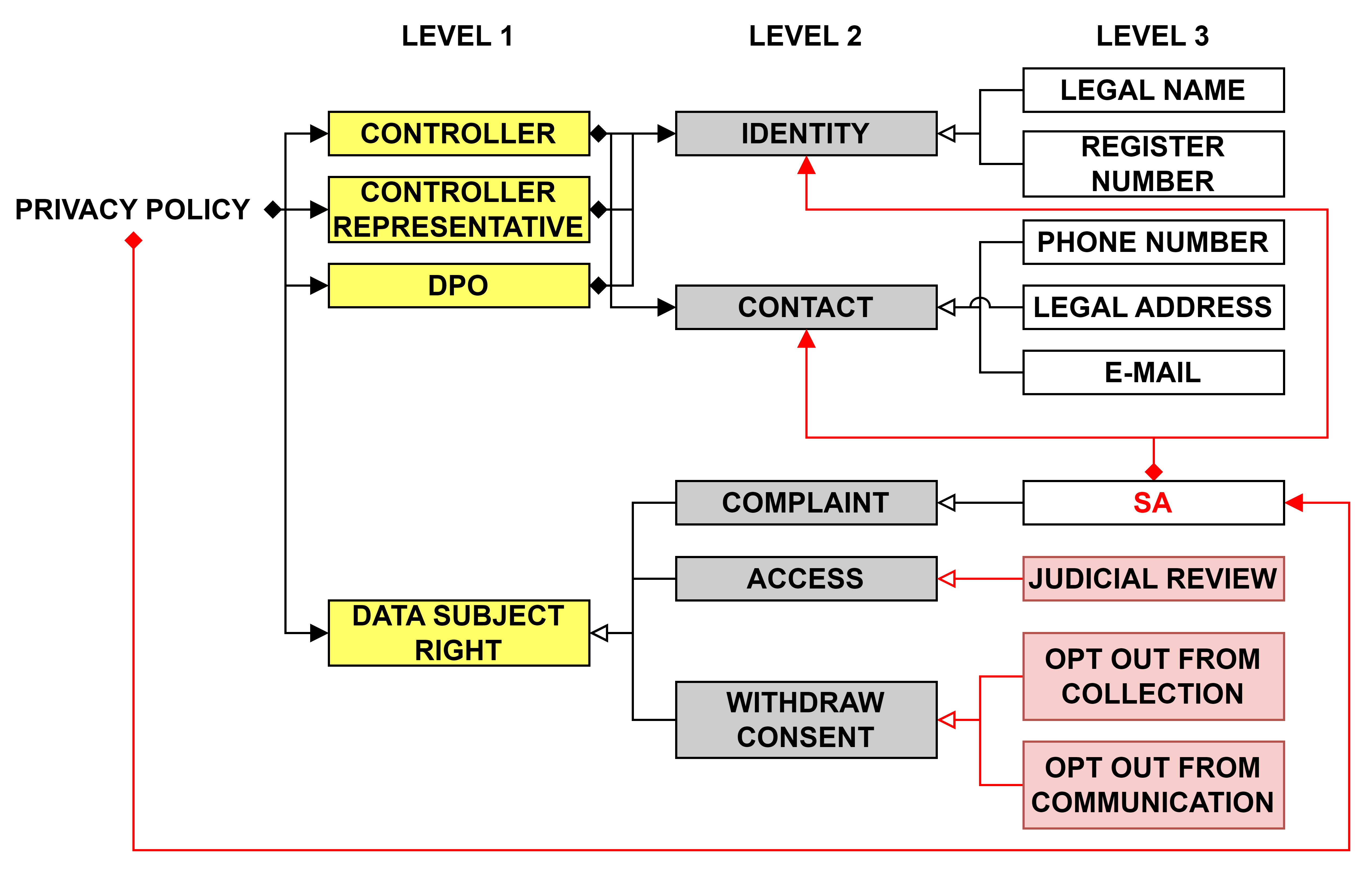}
    \caption{An example of the conceptual model of privacy policy metadata types proposed in \citeauthor{Amaral2021} \cite{Amaral2021, Amaral}. The full conceptual model can be found in \cite{Amaral}. The red boxes and edges added to the original model are the samples of additional metadata types and relationships identified in our study.}
    \label{fig:metadata}
\end{figure}

\newtext{The first step involves collecting privacy policies from existing software applications and identifying and extracting privacy clauses from the narrative statements in those policies. In our study, we focused on online learning applications (OLAs)}, thus we collected the privacy policies of ten popular OLAs from their websites, including Brilliant \cite{Brilliant2020}, Coursera \cite{Coursera}, edX \cite{edX2021}, FutureLearn \cite{FutureLearn2022}, MasterClass \cite{MasterClass2020}, Mindvalley \cite{Mindvalley2021}, Moodle \cite{Moodlea}, SkillShare \cite{Skillshare2019}, Udacity \cite{Udacity2022} and Udemy \cite{Udemy2021}. We then used the metadata types proposed in \cite{Amaral2021} to extract relevant statements in the privacy policies -- referred to as the \emph{privacy clauses}. According to \cite{Amaral2021}, they derived 56 metadata types from the GDPR provisions. These are expected to be present in privacy policies that comply with GDPR (e.g., controller, data subject rights and processing purposes). They also developed a conceptual model to illustrate the hierarchical structure among those metadata types, outlined in Figure \ref{fig:metadata}. This structure defines 113 possible relationships. % which will be later used as categories in our study. 
We extracted 1,081 privacy clauses from the ten privacy policies of the OLAs, summarised in Table \ref{tab:number-of-clauses}. After extracting the privacy clauses from the selected privacy policies, they were classified into 56 categories based on their metadata types. A complete list of metadata types and their associated extracted clauses is included in the replication package \cite{reppkg-privacygen}).
The metadata types are structured into three hierarchical levels: LEVEL 1, LEVEL 2 and LEVEL 3. The following example demonstrates the metadata types and their relationship. 

\begin{table}[ht]
\footnotesize
    \centering
    \caption{Descriptive statistics of the selected OLAs.}
    \label{tab:number-of-clauses}
    \begin{tabular}{@{}llcc@{}}
        \toprule
        \textbf{Privacy policy} & \multicolumn{1}{c}{\textbf{Version}} & \textbf{\#Total clauses}	& \begin{tabular}[c]{@{}c@{}}\textbf{\#Extracted}\\\textbf{clauses}\end{tabular} \\ \midrule
        Brilliant      		& January 2020		& 132		& 113                            \\
        Coursera       		& January 2022		& 246		& 99                            \\
        edX            		& November 2021		& 137		& 74                            \\
        FutureLearn    		& March 2022		& 151		& 108                            \\
        MasterClass    		& January 2020		& 123		& 100                            \\
        MindValley     		& August 2021		& 195		& 80                            \\
        Moodle         		& June 2022			& 85		& 50                            \\
        SkillShare     		& April 2019		& 185		& 117                            \\
        Udacity        		& February 2022		& 145		& 153                            \\
        Udemy          		& December 2021		& 207		& 187                            \\ \midrule
        \multicolumn{2}{c}{\textbf{Total}} 		& \textbf{1,606}	& \textbf{1,081} \\
        \bottomrule
    \end{tabular}
\end{table}

The metadata type \texttt{CONTROLLER} is at the first level, representing a controller who is the main responsible entity controlling the collection and processing of personal data. The \texttt{CONTROLLER} metadata type consists of \texttt{IDENTITY} and \texttt{CONTACT} metadata types at the second level. The first item refers to the identity of the controller, while the latter refers to the contact of the controller. At the third level, the \texttt{IDENTITY} metadata type consists of \texttt{LEGAL NAME} and \texttt{REGISTER NUMBER} metadata types, which means the identity of the controller can be the legal name and register number. This relationship forms two categories in the privacy clause library, which are \texttt{CONTROLLER.IDENTITY.LEGAL NAME} and \texttt{CONTROLLER.IDENTITY.REGISTER NUMBER} respectively. Similarly, the \texttt{CONTACT} metadata type consists of \texttt{PHONE NUMBER}, \texttt{LEGAL ADDRESS} and \texttt{E-MAIL} metadata types, which means the contact of the controller can be phone number, legal address and e-mail. The controller representative and DPO metadata types have the same relationship as the \texttt{CONTROLLER} metadata type.

\subsubsection{Manual Extraction} We manually reviewed each statement in the selected privacy policies and considered whether it matches the metadata types. For instance, a statement in Coursera's privacy policy states that ``\emph{Coursera, Inc. is the data controller of the personal information we collect about you}''. This statement indicates the legal name of the controller, thus we extracted that statement and placed it into \texttt{CONTROLLER.IDENTITY.LEGAL NAME} category in our privacy clause library. Another example is the statement from Moodle's privacy policy stating that ``If you have any questions about our privacy notice, please contact us by email at privacy@moodle.com''. This statement provides the e-mail contact of the controller, thus it was extracted and placed into \texttt{CONTROLLER.CONTACT.E-MAIL} category in our privacy clause library.

\subsubsection{Missing categories} We note that no statements related to the following ten categories was observed in any of the selected privacy policies: the register number of the controller representative (\texttt{CONTROLLER REPRESENTATIVE.IDENTITY.REGISTER~NUMBER}), the phone number of the controller (\texttt{CONTROLLER.CONTACT.PHONE NUMBER}), the phone number and register number of data protection officer (DPO) (\texttt{DPO.CONTACT.PHONE NUMBER} and \texttt{DPO.IDENTITY.REGISTER~NUMBER}), the categories of personal data accessed from public (\texttt{PD CATEGORY.TYPE.PUBLICLY}) or from third party sources (\texttt{PD CATEGORY.TYPE.PUBLICLY}), the personal data collected indirectly from public (\texttt{PD~ORIGIN.INDIRECT.PUBLICLY}) and the transfer of personal data to other countries on the basis of European Commission's adequacy decisions to other countries, sectors or territories outside Europe (\texttt{TRANSFER OUTSIDE EUROPE.ADEQUACY DECISION.\{COUNTRY, SECTOR and TERRITORY\}}). However, the absence of these metadata types cannot lead to the conclusion that the privacy policy is incomplete. We must cross-check with the completeness criteria stated in \cite{Amaral2021}. For example, the first criterion (C1) states that the \texttt{CONTACT} can either be a legal address, phone number, email or all of them. Similarly, the second criterion (C2) states that the \texttt{IDENTITY} can either be a legal name, register number or both. Therefore, at least one metadata type under \texttt{CONTACT} and \texttt{IDENTITY} must be present in a privacy policy to make it complete in this regard.

During the extraction process, we noticed that some common clauses were not covered by the existing metadata types. There are 20 common clauses explaining how users can opt out of personal data collection and communications (e.g., direct marketing emails or newsletters) and six common clauses providing the contact of supervisory authority (SA). These two sets of clauses should be communicated to users via privacy policies. Thus, we added two additional metadata types, which are \texttt{OPT OUT FROM COLLECTION} and \texttt{OPT OUT FROM COMMUNICATION} to cover those statements (see the red boxes in Figure \ref{fig:metadata}). Both metadata types are the descendants of \texttt{WITHDRAW CONSENT} under \texttt{DATA SUBJECT RIGHT}. This creates two relationships: (1) users have the right to opt out from personal data collection \texttt{(DATA SUBJECT RIGHT.OPT OUT FROM COLLECTION)}; and (2) users have the right to opt out from communications \texttt{DATA SUBJECT RIGHT.OPT OUT.COMMUNICATION}. For the \texttt{SA} metadata type (highlighted in the red text in Figure \ref{fig:metadata}), we added the additional link from the \texttt{SA} to \texttt{IDENTITY} and \texttt{CONTACT} metadata types to cover the text that discusses the information of SA. This makes 55 categories in the privacy clause library.

In addition, we have identified an additional individual right from the clause which is the right to judicial review. This right allows individuals to ask organisations to provide reasons when their request under the right of access has been refused. This right is dependent on the right of access, thus it was added under \texttt{DATA SUBJECT RIGHT.ACCESS}. Our privacy clause library ends up with 56 categories.

%% file: sections/4-clauses-refinement.tex
\subsection{Privacy clause refinement} \label{sec:clause-refinement}

Privacy clauses extracted from different privacy policies can be similar or duplicate. The duplicate clauses in our study refer to multiple clauses with the same meaning. In this step, we perform three actions: (1) retain privacy clauses; (2) rephrase privacy clauses; and (3) remove duplicate privacy clauses. As we placed all the extracted privacy clauses into categories in the previous step, we analysed and refined the privacy clauses that are in the same categories. The privacy clauses refinement is performed based on the following rules (R):

\textbf{R0 - Retain:} If a clause is a stand-alone clause and is not duplicated with any other clauses, we retain the original clause in the library. For example, the data subject right category contains a clause that informs the rights of a resident of the European Economic Area. We retained this clause in the library.
	
\textbf{R1 - Rephrase:} For some categories, we rephrase the privacy clauses to make them clearer by leaving out unnecessary words and putting them in a specific template. Six cases applied under this rule are:
	
\begin{itemize}[leftmargin=*]

    \item \textbf{R1.1:} For the \texttt{DATA SUBJECT RIGHT} metadata type (DATA SUBJECT RIGHT.*\footnote{* refers to the right to access, rectification, restriction, complaint, erasure, object to processing, data portability, withdraw consent and judicial review.}), we rephrase the original privacy clause that explains the meaning of that right into \emph{``You have the right/ability to ...''}, followed by the identified data subject rights. For example, the original privacy clause in the \texttt{DATA SUBJECT RIGHT.RECTIFICATION} states that \emph{``(GDPR) right of rectification -- you have a right to correct data that we hold about you that is inaccurate or incomplete.''}. We rephrased this clause to \emph{``You have the right to correct data that we hold about you that is inaccurate or incomplete.''}.
    
    \item \textbf{R1.2:} For the \texttt{REGISTRATION NUMBER} metadata type, we rephrase as \emph{``Our registration number is ...''}, followed by the identified business registration number. For example, the original clause is \emph{``registered number 8324083''}, we rephrased this clause to \emph{``Our registration number is [CONTROLLER'S REGISTER NUMBER]''}. We introduce the use of a \emph{\texttt{placeholder}} here. The placeholder is a temporary field that can be filled with different values. It is denoted by square brackets \textit{\texttt{[p]}}, where \emph{p} is a value identified in the original privacy clause. A privacy clause that conveys the same message but has varied multiple values across different privacy policies (e.g., email, address) requires the placeholder.    
    
    \item \textbf{R1.3:} When a privacy clause contains specific names of parties, applications or third-party service providers, we replace those specific words with general words. For example, the original clause in \texttt{PD ORIGIN.INDIRECT.THIRD~PARTY} states that ``Partner sites providing Content Offering-related tools and services to Coursera users may collect nonfinancial individual-level user data regarding the individual’s use of that partner site while the partner sites provide those services to Coursera.'' is rephrased as \emph{``Our partners may collect your personal information for specific purposes while those partners provide the services to us.''}. 
    
    \item \textbf{R1.4:} For the \texttt{LEGAL ADDRESS} metadata type, we rephrase as \emph{``You can contact us by mail at [LEGAL ADDRESS]}. As the \texttt{LEGAL ADDRESS} is in the third level of the hierarchy in the conceptual model, the value of \texttt{LEGAL ADDRESS} can be the legal address of the controller \texttt{[CONTROLLER.CONTACT.LEGAL ADDRESS]}, controller representative \texttt{[CONTROLLER REPRESENTATIVE.CONTACT.LEGAL ADDRESS]}, DPO \texttt{[DPO.CONTACT.LEGAL ADDRESS]} or SA \texttt{[SA.CONTACT.LEGAL ADDRESS]}. For example, the original clause provides the legal address of the DPO as ``Our independent Data Protection Officer is: Data Compliance Europe Ltd., 12 City Gate, Lower Bridge Street, Dublin 8, Ireland.''. The value of the address is assigned to the placeholder [DPO.CONTACT.LEGAL ADDRESS]. We then rephrased the clause to \emph{``You can contact us by mail at Data Compliance Europe Ltd., 12 City Gate, Lower Bridge Street, Dublin 8, Ireland.''}.
    
    \item \textbf{R1.5:} For the \texttt{DPO.IDENTITY.LEGAL NAME} metadata type, we rewrite as \emph{``We have appointed [DPO.IDENTITY.LEGAL NAME]} as our Data Protection Officer (DPO). For example, the original clause states that \emph{``Udacity EMEA Holdings Ltd. has appointed Bird \& Bird DPO Services SRL as a Data Protection Officer (DPO)''}, we rewrote it to \emph{``We have appointed Bird \& Bird DPO Services SRL as our Data Protection Officer (DPO)''}. 
    
    \item \textbf{R1.6:} For the privacy clauses that are incomplete or require a subject, we add \emph{``We''}, followed by the identified action and object to the rephrased clauses. For example, the privacy clause identified from the Coursera privacy policy states that \emph{our use of your personal information is necessary for complying with our legal obligations}, the clause was rewritten as \emph{``We process your personal information when it is necessary to comply with legal obligations.''}.

\end{itemize}

\textbf{R2 - Remove:} To remove duplicates, we consider privacy clauses that have the same explanation but are written differently. We then select one clause that we judge best describes and covers all the required components of the original clauses. We need to consider the following scenarios: 
	
\begin{itemize}[leftmargin=*]
    
    \item \textbf{R2.1:} If the original clauses have the same meaning but have multiple values across the different privacy policies, we replace those values with the placeholder in the representative clause. For example, there are 84 clauses in the \texttt{PD ORIGIN.DIRECT} category describing from which activities a system directly collects personal data from users. We selected the representative clause, which is ``We directly collect personal information from you when you interact with us. This can be through our websites, over the phone, in person, including, without limitation, when you: [DIRECT SOURCES].''. Other clauses are represented by the placeholder [DIRECT SOURCES].	
    
    \item \textbf{R2.2:} If the original clauses have the same explanation but are written differently, we select one clause to represent all the original clauses. These clauses do not require a placeholder. The following example demonstrates this rule. In the \texttt{PD PROVISION OBLIGED} category, one of the five original clauses was selected to inform users about the consequences if they choose not to provide specific personal data for processing. The representative clause is ``If you choose not to provide certain information required to provide you with various products and services offered on the site, then you may not be able to use some or all parts of our services that require such information.''
    
    \item \textbf{R2.3:} If the original clauses have the same meaning but are written differently and contain additional conditions, we select one of the clauses as a principal clause. We then take those additional conditions as subordinate clauses. These conditions are used to generate questions for the rule-based system in Section \ref{sec:rule-based}. For instance, a set of clauses in the \texttt{CONTROLLER.CONTACT.E-MAIL} category informs users of the email address to contact the controller if they have any questions or concerns about the privacy policy. However, several clauses contain a condition that they only apply to European Union or Swiss citizens. Therefore, we prepared a question to filter this condition in the rule-based system asking \emph{``Are users of your system resided both in AND outside of the European Economic Area or Switzerland?''}
    
    \item \textbf{R2.4:} If multiple original clauses describing the same content contain multiple actions (e.g., collect, process and maintain), we combine every action in a refined clause. For instance, multiple clauses are classified into \texttt{DATA SUBJECT RIGHT.PORTABILITY} category individually states that users should be able to easily download personal data, request a copy of their personal data and request the transfer of their personal data. Thus, we combine these actions and refine them as \emph{``You have the right to obtain, download, request a copy or have your personal information transferred to another organisation.''}.

    \item \textbf{R2.5:} If multiple clauses are written exactly the same, we keep only one clause. For example, two clauses extracted from Brilliant and SkillShare privacy policies were written exactly the same as  \emph{``We will only process your Personal Data if we have a lawful basis for doing so.''}. Both clauses were categorised into \texttt{LEGAL BASIS} category.

\end{itemize}

\textbf{R3 - Retain + Replace:} If a clause is not duplicated but requires a placeholder, we retain that clause and replace the placeholder where it may have varied multiple values. For instance, a clause extracted from Brilliant's privacy policy states, ``By using the Services, you acknowledge that any Personal Data about you, provided by you, is being provided to Brilliant in the U.S. and will be hosted on U.S. servers.''. It was categorised into the \texttt{PD ORIGIN.DIRECT} category. The clause is not duplicated with other clauses in the \texttt{PD ORIGIN.DIRECT} category; however, the text \emph{Brilliant in the U.S.} represents the controller in the U.S., which may have a different value in other privacy policies. Thus, we replace that text with the placeholder \texttt{[US CONTROLLER'S NAME]}.

We note that some of the privacy clauses can be refined by one or more rules. The following example demonstrates this scenario. Two privacy clauses in the \texttt{DATA SUBJECT RIGHT.WITHDRAW CONSENT} metadata type were extracted from Coursera and Moodle privacy policies. The first privacy clause from Coursera states that \emph{``If we rely on your consent for us to use your personal information in a particular way, but you later change your mind, you may withdraw your consent by visiting your profile page and clicking the box to remove consent and we will stop doing so.''}, while the second clause from Moodle says \emph{``You can withdraw your consent at any time either by `selecting delete my data' within the specific service or by request to privacy@moodle.com''}. When we performed the refinement, we found that both clauses explained to the users how to withdraw consent from their applications. We applied \textbf{R1.3} and \textbf{R2.2} to refine these clauses. We finally got \emph{``You can withdraw your consent at any time by following our instructions presented on the site.''} as the refined privacy clause.

%% file: sections/5-rule-basedQA.tex
\section{Privacy policy generation} \label{sec:rule-based}

We have developed an interactive rule-based system for privacy policy generation. Its target users are privacy policymakers i.e., OLA developers in this paper. The privacy policy generation consists of two parts -- interactive privacy policy engine (Engine) and privacy policy generator (Generator). The \texttt{Engine} asks a privacy policymaker a series of questions. The questions are selected based on the answers and rules set inside the system. The \texttt{Generator} uses these to generate a GDPR-compliant privacy policy.

\subsection{Engine}

The Engine part consists of 167 questions, categorised into 10 sections as shown in Table \ref{tab:RQA-sec}. Each question in the Engine system contains the following eight fields: question number (Q\#), question, type of question, referred question, possible answer(s), question flow, placeholder and associated privacy clause(s). The details of those fields are described below:

%\begin{itemize}[leftmargin=*]
\textbf{Question number (Q\#):} a unique identifier and also tracking the number of questions in the system. The question number begins with ``Q", followed by a number (e.g., Q1 and Q166).
	
\textbf{Question:} the description of a question (e.g., ``Who is the controller of personal data collection and processing?'').
 
 \begin{table}[ht]
 \footnotesize
	\centering
	\caption{The details of sections and number of questions in the PPGen system.}
	\label{tab:RQA-sec}
	\begin{tabular}{@{}clc@{}}
		\toprule
		\textbf{Section} & \multicolumn{1}{c}{\textbf{Section name}}  & \textbf{\# of questions} \\ \midrule
		A                & GDPR compliant survey           & 1                        \\
		B                & Contact information             & 27                       \\
		C                & Data subject rights             & 64                       \\
		D                & Personal data storage           & 12                       \\
		E                & Legal basis                     & 23                       \\
		F                & Security                        & 5                        \\
		G                & Personal data provision obliged & 2                        \\
		H                & Processing purposes             & 23                       \\
		I                & Transfer                        & 6                        \\
		J                & Children's data                 & 4                        \\ \midrule
		\multicolumn{2}{c}{\textbf{Total}}                 & \textbf{167}        	  \\ \bottomrule
	\end{tabular}
\end{table}
	
\textbf{Type of question:} specifies the type of a question. There are three types of questions in the PPGen system: 1) closed questions (denoted as \emph{BOOL}); 2) open questions (denoted as \emph{INFO}); and 3) multiple choice questions (denoted as \emph{MTPC}). The closed questions are asked to get either \emph{\textbf{yes}} or \emph{\textbf{no}} answer (e.g., \emph{``Do you have a controller representative?''}). The open questions are asked to seek for detailed and specific answers (e.g., \emph{``Who is the controller of personal data collection and processing?''}). The multiple choice questions provide multiple options to be selected as answers (e.g., \emph{``Which types of personal data does your system collect from users?''}).
	
\textbf{Referred question:} refers to the Q\# with its possible answers in the next field.
	
\textbf{Possible answer(s):} presents all the possible answer(s) based on the type of question. The possible answers for closed questions are [YES] and [NO]. The possible answers for open questions are placeholders that can store values answered by privacy policymakers. Lastly, the possible answers for the multiple choice questions are a list of options. Those selected options are stored in a specific placeholder (e.g., [PD TIME STORED CRITERIA]).
	
\textbf{Question flow:} identifies the next question to be answered.
	
\textbf{Placeholder:} stores open answer question values.
	
\textbf{Associated privacy clause(s):} contains suggested privacy clause(s) that will be added to a privacy policy generated by the PPGen system. The privacy clauses are the clauses that were extracted from Section \ref{sec:clause-development}, and are associated with the possible answer field. In addition, we have added 58 non-compliant and 14 warning clauses that will be placed in the generated privacy policy where needed. Non-compliant clauses are issued when the information that must be included in the privacy policy is missing. These clauses inform the privacy policymaker of the specific information that makes this privacy policy non-compliant with GDPR. The warning clauses are issued when the system detects any inconsistent answer or uncertain statement during the privacy policy generation process. The privacy policymaker needs to review those warning clauses before publishing the privacy policy. %The non-compliant clauses are highlight in \textcolor{red}{\emph{red}} in the privacy policy. The warning clauses are highlighted in \textcolor{yellow}{\emph{yellow}} in the privacy policy.

To form a question, we began with the privacy clauses containing in the library discussed in Section \ref{sec:clause-development}. For example, a privacy clause in the \texttt{CONTROLLER.IDENTITY.LEGAL NAME} in the library states that \emph{``[CONTROLLER'S LEGAL~NAME] is the data controller of the personal information we collect about you.''}, we then converted this clause into a question as \emph{``Who is the controller of personal data collection and processing?''} (see Table \ref{tab:RQA-sample}). This question aims to ask for the controller's legal name, hence it is considered as the open question type. With this type of question, there is no possible answer indicated in the answer selection field. However, the placeholder must receive the value of \emph{[CONTROLLER'S LEGAL NAME]} inputted by a privacy policymaker. Once the privacy policymaker inputs the legal name of their controller, the value is recorded in [CONTROLLER'S LEGAL NAME] placeholder. The associated privacy clauses [C2] ``[CONTROLLER'S LEGAL NAME] is the data controller of the personal information we collect about you.'' \textbf{AND} [C3] ``In this privacy policy, references to ``we'' or ``us'' are to [CONTROLLER'S LEGAL NAME].'' are placed in the target privacy policy generated by the system.

\input{sections/RQA-sample}

\subsection{Generator}

To generate privacy policies, we have developed a template with the following ten sections:
    \begin{itemize}[leftmargin=*]
    \item \textbf{Section 1:} This section provides a brief description of a controller who collects and processes the personal data of a system.
    
    \item \textbf{Section 2:} This section defines personal data, purposes of personal data collection, categories of personal data collected and how personal data is collected in the system.
    
    \item \textbf{Section 3:} This section specifies how the system and its service providers/third parties use personal data.
    
    \item \textbf{Section 4:} This section outlines the purposes of personal data storage and retention. It also explains how the system manages personal data when it is kept in the system.
    
    \item \textbf{Section 5:} This section informs the users of the system of their rights and how they can execute those rights.
    
    \item \textbf{Section 6:} This section provides information regarding the transfer of personal data outside Europe. It also lists the recipients of the transferred personal data and the regulations that are used to govern the transfer.
    
    \item \textbf{Section 7:} This section mentions the collection and use of children's data in the system. If the system collects children's data without valid parental consent, it provides a guideline for parents to report this issue to the system.
    
    \item \textbf{Section 8:} This section explains a process for filing and submitting a complaint to the controller or supervisory authority. It also explains how the controller handles the complaints.
    
    \item \textbf{Section 9:} This section specifies technical and organisational security measures used to protect personal data in the system. This includes the notification of data breaches to the affected users.
    
    \item \textbf{Section 10:} This section provides the contact details of the data controller, controller representative and DPO.
\end{itemize}

To form a privacy policy, we populate the template with section headings and text to guide the readers. We selected answer placeholders from Engine questions and placed them into the relevant sections in the template. We designed a notation to represent a question in the template as: \smallskip

\small \hbox{\hss [Q\#-Type.PA$\rightarrow$[Placeholder]$\rightarrow$[Associated privacy clause(s)]$\rightarrow$QF] \hss}\smallskip

This notation determines where an output for each question should be placed in the template. The following example demonstrates this step. We placed Q3 - Q26 into Section 10 in the template as they relate to the contact information of the data controller, controller representative and DPO. For Q3, it asks the privacy policymaker to provide a legal address of the controller (see Table \ref{tab:RQA-sample}). This question is an open question (INFO) that does not have suggested possible answer(s). It will collect the legal address inputted by the privacy policymaker and record it in the [CONTROLLER'S LEGAL ADDRESS] placeholder. There is no associated privacy clause identified for this question, hence the \emph{[Associated privacy clause(s)]} is empty. The question flow indicates that the next question after Q3 is Q4. We thus represent this question in the privacy policy template as [Q3-INFO$\rightarrow$[CONTROLLER'S LEGAL ADDRESS]$\rightarrow$Q4]. This question only collects the controller's legal address but does not provide any outputs in the template.

The following examples show how we derived the notations for Q88 and Q89. Both questions involve personal data storage and were placed in Section 4 in the privacy policy template. Q88 is a closed question (BOOL) that asks if the system stores personal data. This question has two possible answers: YES or NO. If the answer is YES, the next question will be Q89, otherwise, the next question will be Q93. The question does not have the associated privacy clauses, thus here are the two notations derived from Q88: [Q88-BOOL.YES$\rightarrow$Q89] and [Q88-BOOL.NO$\rightarrow$Q93]. Q89 is a multiple choice question (MTPC) that allows the privacy policymaker to specify the criteria for storing personal data in the system. Those criteria will be recorded in the [PD TIME STORED CRITERIA] placeholder. The next question after Q89 is Q90. The notation for Q89 is [Q89-MTPC$\rightarrow$[PD TIME STORED CRITERIA]$\rightarrow$Q90].

Another example demonstrates the notation of a question with associated privacy clause(s). Q166 is an open question (INFO) that asks for the controller's registration number. This question was placed in the first section of the privacy policy template. The controller's registration number inputted by the privacy policymaker will be recorded in the [CONTROLLER'S REGISTER NUMBER] placeholder. Then, clause [C4] ``Our registration number is [CONTROLLER'S REGISTER NUMBER] must be presented in the generated privacy policy. Hence, the notation of Q166 in the privacy policy template is represented as [Q166-INFO$\rightarrow$[CONTROLLER'S REGISTER NUMBER]$\rightarrow$C4$\rightarrow$Q3].

\subsection{Implementation}

\begin{figure}[ht]
    \centering
    \includegraphics[width=1.0\linewidth]{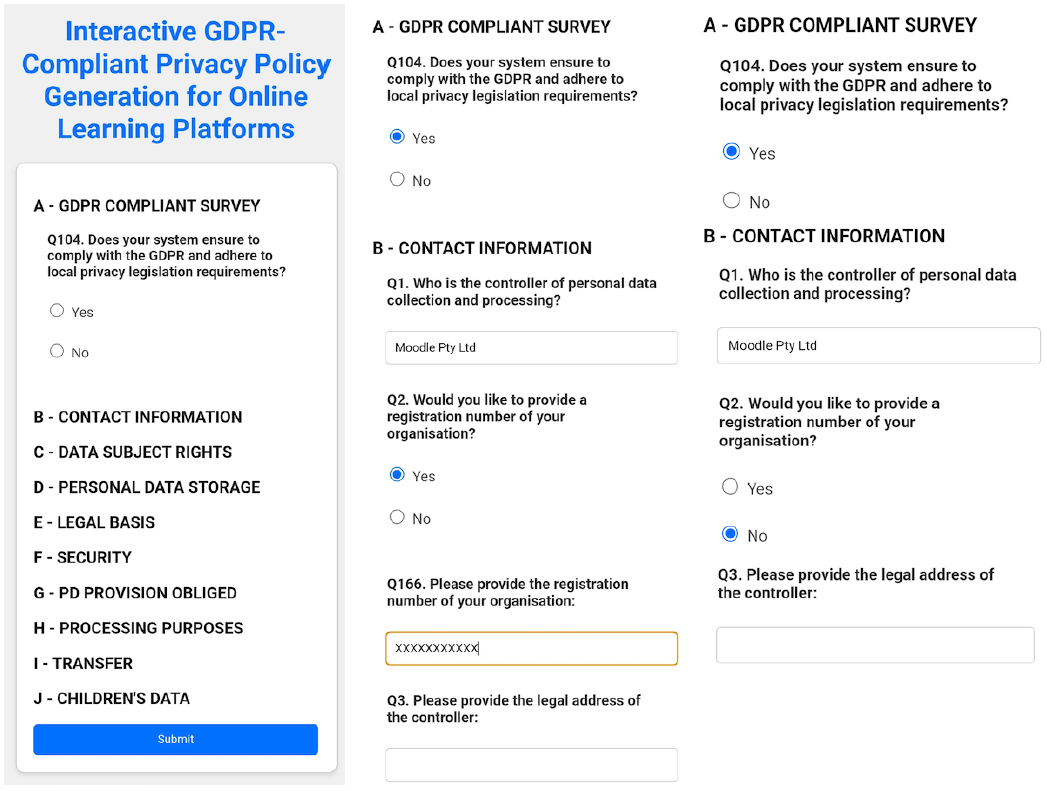}
    \caption{Screenshots of our PPGen tool in use}
    \label{fig:prototype}
\end{figure}

We have implemented PPGen as a web-based application and made it available at \emph{\url{https://ppgen-tool.netlify.app}}. We used Netlify, linked with Github, to develop and deploy our tool. The tool was written using HTML, CSS and JavaScript. Figure \ref{fig:prototype} shows a snapshot of the tool. PPGen first shows its landing page, which outlines all the sections that a software developer must complete. It starts by asking the developer the first default question (i.e., Q104). Once the developer completes answering the first question, the response will be used to determine the next question, which will be presented to the developer automatically. For example, if an OLA developer answers \emph{Yes} in Q104, the tool then shows Q1. As Q1 is an open question, it only allows the developer to fill in the information in the provided text field. After the developer completes Q1, the tool shows Q2. The developer can answer \emph{Yes} or \emph{No} in Q2, depending on their software application. If the developer answers \emph{Yes}, Q166 will be presented, and the developer must complete Q166 before proceeding to Q3. On the other hand, if \emph{No} is selected, Q3 will be presented. Our tool is interactive and provides an incremental privacy policy to developers. When the developer completes all the questions, they can press a \emph{Submit} button to view an entire list of customised privacy clauses for their privacy policy. The developer can also modify their answers and resubmit to regenerate the privacy clauses. We have also provided a \emph{short demo video} at \emph{\url{https://bit.ly/PPGen-demo}}.

%% file: sections/RQA-sample.tex
\begin{table*}[!htbp]
\footnotesize
    \centering
    \caption{Sample of questions and their associated fields in the PPGen system.}
    \label{tab:RQA-sample}
        \resizebox{7in}{!}{%
        \footnotesize
    \begin{tabular}{lllp{0.5cm}p{0.5cm}p{0.5cm}lp{7.5cm}}
        \hline
        \multicolumn{1}{c}{\textbf{Q\#}} &
        \multicolumn{1}{c}{\textbf{Question}} &
        \multicolumn{1}{c}{\textbf{Type}} &
        \multicolumn{1}{c}{\textbf{RQ}} &
        \multicolumn{1}{c}{\textbf{PA}} &
        \multicolumn{1}{c}{\textbf{QF}} &
        \multicolumn{1}{c}{\textbf{Placeholder}} &
        \multicolumn{1}{c}{\textbf{Associated privacy clause(s)}} \\ \hline
        \multicolumn{8}{l}{\textbf{SECTION B - CONTACT INFORMATION}} \\ \hline
        \multicolumn{1}{c}{Q1} &
        \begin{tabular}[t]{@{}l@{}}Who is the controller of \\ personal data collection \\ and processing?\end{tabular} &
        INFO &
        &
        &
        Q2 &
        \begin{tabular}[t]{@{}l@{}}{[}CONTROLLER'S \\ LEGAL NAME{]}\end{tabular} &
        \begin{tabular}[t]{@{}l@{}}{[}C2{]} {[}CONTROLLER'S LEGAL NAME{]} is the data controller of \\the personal information we collect about you.\\ AND\\ {[}C3{]} In this privacy policy, references to "we" or "us" are to \\{[}CONTROLLER'S LEGAL NAME{]}.\end{tabular} \\ \hline
        \multicolumn{1}{c}{Q2} &
        \begin{tabular}[t]{@{}l@{}}Would you like to provide \\ a registration number of \\ your organisation?\end{tabular} &
        BOOL &
        Q2 Q2 &
        {[}YES{]} {[}NO{]} &
        Q166 Q3 &
        &
        \\ \hline
        \multicolumn{1}{c}{Q166} &
        \begin{tabular}[t]{@{}l@{}}Please provide a registration \\ number of your organisation.\end{tabular} &
        INFO &
        &
        &
        Q3 &
        \begin{tabular}[t]{@{}l@{}}{[}CONTROLLER'S \\ REGISTER NUMBER{]}\end{tabular} &
        \begin{tabular}[t]{@{}l@{}}{[}C4{]} Our registration number is\\ {[}CONTROLLER'S REGISTER NUMBER{]}.\end{tabular} \\ \hline
        \multicolumn{1}{c}{Q3} &
        \begin{tabular}[t]{@{}l@{}}Please provide the legal address\\of the controller.\end{tabular} &
        INFO &
        &
        &
        Q4 &
        \begin{tabular}[t]{@{}l@{}}{[}CONTROLLER'S \\ LEGAL ADDRESS{]}\end{tabular} &
        \\ \hline
        \multicolumn{1}{c}{Q4} &
        \begin{tabular}[t]{@{}l@{}}Please provide the email address\\of the controller.\end{tabular} &
        INFO &
        &
        &
        Q5 &
        \begin{tabular}[t]{@{}l@{}}{[}CONTROLLER'S \\ EMAIL {]}\end{tabular} &
        \\ \hline
        \multicolumn{1}{c}{Q5} &
        \begin{tabular}[t]{@{}l@{}}Does your system allow users \\to contact the controller if \\they have any questions or \\concerns related to their rights, \\concerns, comments or complaints\\ about this privacy policy?\end{tabular} &
        BOOL &
        Q5 Q5& 
        {[}YES{]} {[}NO{]}&
        Q6 Q6&
        &
        \begin{tabular}[t]{@{}l@{}}{[}C5{]} If you have any questions about this privacy policy or \\our data practices generally, please contact us using the following\\ information: {[}CONTROLLER'S LEGAL NAME{]} \\ {[}CONTROLLER'S LEGAL ADDRESS{]}. \\ AND\\ {[}C6{]} or at {[}CONTROLLER'S EMAIL{]}.\end{tabular} \\ \hline
        \multicolumn{8}{l}{\textbf{SECTION D - PERSONAL DATA STORAGE}} \\ \hline
        \multicolumn{1}{c}{Q88} &
        \begin{tabular}[t]{@{}l@{}}Does your system store \\ personal data?\end{tabular} &
        BOOL &
        Q88 Q88&
        {[}YES{]} {[}NO{]}&
        Q89 Q93&
        &
        \\ \hline
        \multicolumn{1}{c}{Q89} &
        \begin{tabular}[t]{@{}l@{}}From the list shown below, \\ please select the storage criteria \\ that apply to your system.\end{tabular} &
        MTPC &
        &
        &
        Q90 &
        \begin{tabular}[t]{@{}l@{}}{[}PD TIME STORED \\ CRITERIA{]}\end{tabular} &
        \begin{tabular}[t]{@{}l@{}}Your account is active or as needed to provide you with services\\To serve the purposes which it was collected\\To maintain a record of your transactions for financial reporting,\\ audit, and compliance purposes \end{tabular} \\
        &
        &
        &
        &
        &
        &
        &
        To resolve disputes \\
        &
        &
        &
        &
        &
        &
        &
        To comply with legal obligations \\
        &
        &
        &
        &
        &
        &
        &
        \begin{tabular}[t]{@{}l@{}}To establish, exercise, or defend our legal rights\end{tabular} \\
        &
        &
        &
        &
        &
        &
        &
        \begin{tabular}[t]{@{}l@{}}Permitted or required by applicable law, rule or regulation\end{tabular} \\
        &
        &
        &
        &
        &
        &
        &
        \begin{tabular}[t]{@{}l@{}}Required by specific sector requirements and agreed practices\end{tabular} \\
        &
        &
        &
        &
        &
        &
        &
        \begin{tabular}[t]{@{}l@{}}Required for income tax and audit purposes\end{tabular}  \\ \hline
        \multicolumn{1}{c}{Q90} &
        \begin{tabular}[t]{@{}l@{}}Does your system store \\personal data for other purposes\\ apart from the list above?\end{tabular} &
        BOOL &
        Q90 Q90&
        {[}YES{]} {[}NO{]} &
        Q91 Q92 &
        &
        \\ \hline
    \end{tabular}%
    }
    {\parbox{15.5cm}{\footnotesize Type: type of question, RQ: referred question, PA: potential answer, QF: question flow.}}
\end{table*}

%% file: sections/6-evaluation.tex
\section{Evaluation} \label{sec:evaluation}

\subsection{Research questions and experimental settings}

This evaluation aimed to answer the following research questions:

\begin{itemize}
    \item \textbf{RQ1.} (PPGen vs. existing privacy policy generators) \emph{How does our approach compare to existing privacy policy generators in terms of readability, completeness, and coverage of the generated privacy policy?} \\

    To address this question, we compare the privacy policy generated by our PPGen system to four privacy policies. These are generated by Moodle itself (June 2022 version) and three online privacy policy generator websites, PrivacyPolicies \cite{PrivacyPolicies2022}, Termly \cite{Termly2022} and TermsFeed \cite{TermsFeed2023}. These privacy policy generators were ranked as the top five of the best privacy policy generators of 2023 \cite{Digital}. All allow users to create documents that comply with well-known data protection regulations and laws such as GDPR \cite{OfficeJournaloftheEuropeanUnion;2016} and CCPA \cite{StateofCaliforniaDepartmentofJustice2018}. Some cover specific regulations like the California Privacy Rights Act (CPRA) \cite{CPRA2020}, Children's Online Privacy Protection Act (COPPA) \cite{COPPA2000}, Personal Information Protection and Electronic Documents Act (PIPEDA) \cite{PIPEDA2019} and ePrivacy Regulation \cite{ePrivacy2017}. We used the example of Moodle to generate the privacy policies from these three policy generators with their free standard plans. We followed the steps and answered all the questions asked in the policy generators. All policies are available in our replication package \cite{reppkg-privacygen}. \\
    
    \item \textbf{RQ2.} (PPGen for a new software application) \emph{Can our approach generate a privacy policy for a new software application that does not yet have a privacy policy?} \\

    \newtext{We evaluate whether our approach can generate a high-quality privacy policy for a new software application that does not yet have a privacy policy. For this evaluation, we used Brightspace \cite{D2L2024, Brightspace}, a learning management system software for online learning and teaching. We note that Brightspace's privacy policy was not involved in the privacy clause extraction process; therefore, it is considered a `new' software application for our purposes. We generated a candidate Brightspace privacy policy using our Engine and Generator (refer to Section \ref{sec:rule-based} for more details). We then compared the original Brightspace privacy policy with the one generated by our approach. Both privacy policies are available in our replication package \cite{reppkg-privacygen}.} \\
    
    \item \textbf{RQ3.} (PPGen vs. ChatGPT) \emph{How does our approach compare to ChatGPT in terms of readability, completeness, and coverage when generating privacy policies?} \\

    \newtext{We assess our approach against a generative AI-based tool. Generative AI is a type of artificial intelligence technology that empowers users to rapidly create new content based on various inputs. These systems are designed to understand and mimic human-like patterns, allowing them to create data, text, images, or other forms of content that were not explicitly provided to them during training. Various models and frameworks have been developed by different organisations and researchers. One of the well-known examples includes the Generative Pre-trained Transformer (GPT) series by OpenAI \cite{OpenAI}. Generative AI-based tools, particularly ChatGPT, have gained immense popularity and attention in recent years \cite{KrystalHu2023, Marr2023}. ChatGPT is powered by the GPT series \cite{ChatGPT}, and has been applied to a range of content generation tasks (e.g., writing product reviews, outlining academic articles, producing different types of reports, and writing and debugging codes). A recent study showed that ChatGPT can enhance productivity and significantly save time and effort in mid-level professional writing tasks \cite{Noy2023}. Hence, we assess the potential of using a Generative AI-based tool in generating privacy policies compared to our approach. We consider ChatGPT with GPT-3.5 architecture as a prominent representative of a generative AI-based tool and use it for evaluation.}

    \newtext{The third setting evaluates the privacy policy generated by our PPGen system against the one generated by ChatGPT. To investigate ChatGPT's ability to generate a GDPR-compliant privacy policy based on a provided company description, we initially provided company information to ChatGPT and tasked it with generating a privacy policy. We used Moodle for the evaluation in this experiment.} \\
    
    \item \textbf{RQ4.} (PPGen with ChatGPT) \emph{Can ChatGPT be integrated into our approach, and how does the performance of PPGen with ChatGPT compare to that of the original PPGen?} \\

    \newtext{This setting investigates ChatGPT's ability to assist privacy policymakers in generating a privacy policy. This setting seeks to examine the efficacy of ChatGPT when employed as an assistant to privacy policymakers for the generation of a privacy policy. In particular, we aim to determine whether ChatGPT can accurately answer the questions typically posed to privacy policymakers. The objective of this approach is to streamline the process, enabling privacy policymakers to delegate the task of answering these questions to ChatGPT and consequently reduce the time required for privacy policy generation. To conduct this investigation, we first provided ChatGPT with the original Moodle privacy policy. This step was undertaken to impart knowledge to ChatGPT about the organisation and its system. We then posed the questions from our Engine to ChatGPT, mimicking the steps employed in our approach for privacy policy generation.}

    \newtext{As outlined in Section \ref{sec:rule-based}, the questions are categorised into three types: closed, open, and multiple-choice questions. To better align with ChatGPT's prompt structure, several modifications were made. For closed questions, the questions were slightly modified by changing the reference from ``your system'' to ``Moodle''. For example, Q32, the original question ``Does your system allow users to request the erasure of their personal data?'' was modified to ``Does Moodle allow users to request the erasure of their personal data?''. Open questions were adjusted by transforming requests such as ``Please provide...'' into wh-questions. For instance, Q3, initially ``Please provide the legal address of the controller,'' was changed to ``What is the legal address of the controller?''. Multiple-choice questions were adapted by combining the strategies mentioned above. For example, Q62, the original question ``Which types of personal data does your system collect from users?'' was modified to ``What personal data does Moodle collect from users?''.}
    
\end{itemize}

\subsection{Evaluation criteria} \label{sec:evaluation-criteria}

\subsubsection{Readability} We use the Flesch Reading Ease (FRE) test to measure the readability. This measure has been used to assess the readability of privacy policies in several areas \cite{Graber2002, McDonald2009, Savla2012, Krumay2020}. The FRE score is a 100-point scale, with 100 being the easiest to read. The document with an FRE score of 60 or higher will be fairly easy for the average adult to read \cite{Flesch1948}. The higher score means that the document is easier to read and understand. The FRE score is computed by:
\begin{equation*}
	206.835 - (1.015\times \text{ASL}) - (84.6\times \text{ASW})
\end{equation*}

\noindent where \emph{ASL} is an average sentence length (number of words divided by number of sentences) and \emph{ASW} is an average number of syllables per word (number of syllables divided by number of words). 

We performed the FRE test using Microsoft Word \cite{Microsoft2022}. We also report several relevant readability metrics, including the number of words, number of sentences, number of words per sentence and reading time to support the readability evaluation. Those metrics, except the reading time, were calculated by Microsoft Word. The reading time was calculated by an online tool available at \url{https://niram.org/read/}.

\subsubsection{Completeness} We adopt 23 GDPR completeness checking criteria for privacy policies proposed in \cite{Amaral2021} to evaluate the completeness of the privacy policies generated in this study. Those criteria were derived from the articles in GDPR. They were formulated as pseudo-code statements and written in the following template [\emph{precondition}], $\langle$\emph{postcondition}$\rangle$. Each pseudo-code statement consists of two parts: 1) a \emph{precondition} and 2) a \emph{postcondition}. The preconditions determine metadata type(s) \emph{or} other GDPR-related conditions proposed by the legal experts that need to be identified in a privacy policy. The postconditions are the identification of the metadata type(s) that are different from those identified in the preconditions. To evaluate the completeness criteria, we use three rating values: \emph{satisfied, unsatisfied and precondition not satisfied}. The privacy policy is considered as \emph{complete} if none of the ``must be identified'' criteria (e.g., criteria C2, C3 and C15 in Table \ref{tab:criteria2-results}) is rated \emph{unsatisfied}.

\subsubsection{Coverage} We use a checklist for evaluating coverage of privacy policies. Our checklist was developed based on recommendations of several public bodies (e.g., Information Commissioner's Office \cite{ICO2023}, GDPR.EU \cite{Wolford2019} and PwC Australia \cite{PwCAustralia2023}). It consists of 39 topics divided into 10 categories: 1) general information of a privacy policy (e.g., effective/last updated date and information about controller), 2) personal data collection, 3) personal data use, 4) personal data storage, 5) personal data transfer, 6) individual rights, 7) children's data, 8) complaint handling, 9) personal data protection, and 10) contact information. We use three rating values to evaluate the coverage, which are \emph{Y}, \emph{N} and \emph{W}. The topic is marked with \emph{Y} if the required information is covered in the privacy policy. The topic marked with \emph{N} implies the absence of the required information in the privacy policy. This makes the privacy policy non-compliant with GDPR. Lastly, the topic is marked with \emph{W} if the topic needs to be reviewed by a privacy policymaker. The full checklist is available in our replication package \cite{reppkg-privacygen}.

\subsection{Results \& Discussion}

\subsubsection{\newtext{RQ1. PPGen vs. existing privacy policy generators}}

\textbf{Readability:} Our RQ1 evaluation results investigating privacy policy readability are reported in Table \ref{tab:criteria1-results}. The original Moodle privacy policy achieved the highest FRE score at 45.5, followed by the Termly generated privacy policy with the FRE score at 39. The PPGen generated privacy policy ranked third together with TermsFeed generated privacy policy with a FRE score of 37.4. The privacy policy generated by PrivacyPolicies achieved the lowest FRE score at 37.3. The original Moodle privacy policy has the most optimal values for the pair of ASL and ASW. The Termly generated privacy policy is the longest privacy policy requiring the longest time to read, whereas the original Moodle policy is best in terms of readability. We note that the privacy policies generated by PrivacyPolicies and TermsFeed are almost identical and thus have similar readability results. \input{sections/criteria1-results}

\textbf{Completeness:} Our evaluation found that \emph{the original Moodle privacy policy is not complete}. Out of 23 criteria, 15 criteria were satisfied, 2 criteria were unsatisfied and 6 criteria were rated as precondition not satisfied. All the completeness checking criteria that do not have the precondition were all satisfied. Those two unsatisfied criteria were caused by the missing information of the controller representative. 
\input{sections/criteria2-results}

Table \ref{tab:criteria2-results} shows a few examples extracted from the completeness evaluation (the full result is provided in the replication package \cite{reppkg-privacygen}). Criterion C2 implies that the legal address, email or phone number of the controller must be identified. The contact information of the controller is presented in every privacy policy. Thus, this criterion is satisfied for all policies. Criterion C3 implies that if the controller is located outside of Europe, the registration number or legal name of the controller's representative must be identified. This information was not explicitly stated in the original Moodle privacy policy, however, the headquarters of Moodle is located in Australia. Hence, we considered Moodle to be located outside Europe. This means the legal name or registration number must be identified. However, we found no information about the controller representative in any privacy policies. Therefore, this criterion is unsatisfactory. 

As shown in Table \ref{tab:criteria2-results}, criterion C6 implies that if the right to lodge a complaint is in privacy policies, then the right to lodge a complaint with a supervisory authority should be identified. As C6 is a ``should be identified'' criterion, the metadata types in this criterion are recommended to be present, but the completeness of privacy policies does not depend on this criterion. We found that only the original Moodle and PPGen generated privacy policies satisfied this criterion. Criterion C15 implies that if personal data is collected indirectly, then source(s) must be identified. The original Moodle policy says that Moodle collects personal data from data subjects themselves and does not acquire personal data from other sources. Thus, this precondition was not satisfied and the postcondition was not evaluated for this policy. This result also applied to other privacy policies since it is the condition identified by Moodle. Similarly, the precondition of criterion C16 was not satisfied as the indirect source of personal data was not present in any privacy policies.

\emph{The PPGen generated privacy policy fully aligned with the original Moodle privacy policy.} It has the same evaluation results for all the completeness checking criteria as the original one. This demonstrates that our approach can generate a high-quality, complete privacy policy. The Termly generated policy mostly aligns with the original Moodle policy. However, there is one more criterion that this policy did not satisfy. It does not provide information on how to lodge a complaint with a supervisory authority. The privacy policies generated by PrivacyPolicies and TermsFeed do not closely align with the original Moodle privacy policy. Both have the same evaluation results on every criterion with 5 satisfied, 8 unsatisfied and 10 preconditions not satisfied. They missed a lot of information about data subject rights (e.g., right to complain, data portability, object to processing, personal data transfer and contact details of DPO).

\textbf{Coverage:} We found the PPGen generated privacy policy covers 34 out of 39 topics in the checklist, based on the information available from Moodle. It covers the same topics that are covered by the original Moodle privacy policy. The original Moodle privacy policy did not cover the information related to personal data collection by partners/third parties/data collection tools, personal data collection by email communication/direct marketing, sharing personal data with third parties outside the scope of this privacy policy, notification of personal data breaches and contact information of controller representative. These five topics were presented as two non-compliances and three warnings in our PPGen generated privacy policy. 

The Termly generated privacy policy covers 30 topics, while the PrivacyPolicies and TermsFeed generated privacy policies cover 22 out of 39 topics. They mostly missed the information related to individual rights (e.g., right to judicial review, right to object to automated processing and right to data portability), complaint handling, notification of breaches and contact information of controller representative and DPO. We observed that these policies include information that was not present in the original Moodle privacy policy. This is due to the fixed templates they used not being customisable for several topics (e.g., collecting personal data by partners/third parties/data collection tools). Hence, there are several major limitations in generating a privacy policy from those existing policy generators.

Our PPGen system can also show non-compliances and warnings in a generated policy. For example, none of the privacy policies explains how the controller handles and communicates to affected data subjects and relevant stakeholders when personal data breaches occur, even in the original Moodle privacy policy. Thus, the notification of data breaches was marked with \emph{N} for all the privacy policies. While this piece of information was not provided in the other privacy policies, the PPGen generated privacy policy presented two non-compliant clauses in the privacy policy, which are \emph{``Your system does not comply with the GDPR - notification of a personal data breach.''} and \emph{``Your system does not comply with the GDPR - not notifying a security breach to affected users within 72 hours.''} to inform the OLA developer of information missing. The warning about the controller representative was given in the PPGen generated policy since Moodle did not include the information about the controller representative in its privacy policy. This information should be reviewed by the OLA developer as the controller representative is required for organisations located outside Europe. This topic was marked with \emph{W} in the checklist. Our full evaluation results on coverage can be found in our replication package~\cite{reppkg-privacygen}.

\begin{tcolorbox}[boxsep=0pt,colback=white,colframe=black!75!black]
  \textbf{PPGen outperformed PrivacyPolicies, Termly and TermsFeed in terms of readability, completeness and coverage. The privacy policy generated by PPGen fully aligned with the original privacy policy, albeit with slightly lower readability.}
\end{tcolorbox}

\hspace{1mm}

\subsubsection{\newtext{RQ2. PPGen for a new software application}}

\newtext{\textbf{Readability:} The evaluation results for the readability of the original Brightspace privacy policy and the one generated by our approach are presented in Table \ref{tab:criteria1-results-ii}. The privacy policy generated by our approach yielded a higher FRE score of 39.3 compared to the original version, which scored 28.9.}
\input{sections/criteria1-results-ii}

\newtext{\textbf{Completeness:} Our evaluation revealed that \emph{the original Brightspace privacy policy is incomplete}. Among the 23 criteria assessed, 16 were satisfied, 5 were unsatisfied, and 2 were rated as a precondition not satisfied. The unsatisfied criteria resulted from missing information regarding individual rights and the obligated statement for personal data provision. Examples of the completeness evaluation are provided in Table \ref{tab:criteria2-results-ii}. These results align with those of the Brightspace privacy policy generated by our approach, similar to the evaluation on Moodle. The reason our approach did not achieve better results is that the privacy policymaker (i.e., the author) did not have access to additional sources of information about Brightspace beyond its original privacy policy to provide answers to the Engine. If sufficient information had been available, our PPGen system would have been able to generate a more comprehensive privacy policy for Brightspace.} 
\input{sections/criteria2-results-ii}

\newtext{\textbf{Coverage:} The Brightspace privacy policy generated by our approach addresses 31 out of 39 topics in the coverage checklist. It covers the same topics that are covered by the original Brightspace privacy policy. The original privacy policy, however, did not cover information on eight topics: personal data provision agreement, special categories of personal data, five individual rights (i.e., the right to judicial review, right to object to processing, right to object to automated processing, right to data portability, and consent withdrawal), and notification of data breaches. Among these eight topics, six were absent in the original privacy policy but were incorporated with non-compliant clauses in the PPGen generated privacy policy. Another topic, complaint handling with a supervisory authority, was accompanied by a warning clause in the PPGen generated privacy policy. The final topic, the right to object to automated processing, was not covered in the privacy policy generated by our approach. This exclusion was due to Brightspace lacking functions that make decisions solely by automated means without any user involvement. Thus, this topic was not included in the PPGen generated privacy policy.}

\newtext{Based on the readability, completeness and coverage results, it is evident that our approach can generate a privacy policy for a new software application. It can produce a privacy policy that is consistent with the one manually written. In addition, it can identify missing topics that impact the compliance of a privacy policy, offering valuable information to privacy policymakers for resolving these issues in the privacy policy generation process.}

\begin{tcolorbox}[boxsep=0pt,colback=white,colframe=black!75!black]
  \textbf{PPGen can be used to generate a high-quality privacy policy in terms of readability, completeness and coverage for new software applications.}
\end{tcolorbox}

\hspace{1mm}

%By using information from the original privacy policy, our approach produces a privacy policy that aligns with the original version. However, it may not achieve a better result due to the privacy policymaker (i.e., an author) did not have other sources of additional information about Brightspace beyond its original privacy policy. This evaluation suggests that our approach can generate a privacy policy for a new software application if a privacy policymaker provides sufficient information in our PPGen system.

\subsubsection{\newtext{RQ3. PPGen vs. ChatGPT}} \newtext{The results indicated that ChatGPT could produce a generic Moodle privacy policy, consisting of nine sections. However, the first version of the policy is very short (399 words). Each section was explained briefly and lacked specific details. Later, we provided additional information to ChatGPT with an existing Moodle privacy policy. We requested ChatGPT to modify and add relevant information to the initially generated policy. This updated version extended to 592 words with 17 sections. This revised version integrated some specific details from the Moodle privacy policy and provided more explicit information (e.g., types of personal information, purposes of use and contact information). The privacy policy can be found in the replication package \cite{reppkg-privacygen}. We then evaluated this privacy policy on readability, completeness and coverage. While it took less time to generate a privacy policy using ChatGPT, we observed that the privacy policy produced by ChatGPT underperformed on all criteria when compared to the Moodle privacy policy generated by our approach.}

\newtext{\textbf{Readability:} The FRE score of the privacy policy generated by ChatGPT is low at 19.3, compared to 37.4 for the Moodle privacy policy generated by our approach (see Table \ref{tab:criteria1-results}). The ChatGPT generated privacy policy obviously lacks specific details in every section. Each section comprises only a few sentences. For example, the policy generated by our approach provides a comprehensive explanation of eight individual rights in a dedicated section, whereas the ChatGPT generated policy contains two sentences on this topic. Similarly, the information regarding children’s data is elaborated in five informative sentences in the policy generated by our approach while it is explained in two brief sentences in the ChatGPT generated policy.}

\newtext{\textbf{Completeness:} The privacy policy generated by ChatGPT satisfied 6 out of 23 completeness criteria, whereas the privacy policy generated by our approach satisfied 15 out of 23 completeness criteria. The ChatGPT generated policy lacks several required topics including individual rights (e.g., right to restrict of processing, right to file a complaint with a supervisory authority and right to data portability). Additionally, it does not provide details regarding the transfer of personal data to Europe and other recipients of personal data. We observed that four completeness criteria (i.e., C15 - C19) were marked as "satisfied but not aligned with the original." This indicates that while the criteria were met, the provided information does not match the original privacy policy. For example, C15 in Table \ref{tab:criteria2-results}, the privacy policy generated by ChatGPT included details about the indirect source from which personal data is collected. However, this information differs from what is presented in Moodle's original privacy policy.}

\newtext{\textbf{Coverage:} The ChatGPT generated privacy policy addresses 20 out of 39 topics. Some examples of completeness criteria are shown in Table \ref{tab:criteria2-results}. The following essential topics were not covered: definitions of personal data, purposes of collection, personal data collection through third parties/data collection tools/email communication methods, sharing with third parties, purposes of storage, storage and retention methods, personal data transfer, certain individual rights (e.g., the right to judicial review and the right to object to automated processing), consent withdrawal, measures to protect personal data, notification of personal data breaches, and the contact information of the controller representative. Furthermore, some explanations of specific topics are unclear. For instance, the ChatGPT generated privacy policy includes the statement: ``We prioritize protecting children's personal data. Contact us at privacy@moodle.com for concerns about children's data.'' However, it fails to explicitly state that it does not collect children’s data, as clarified in the original Moodle privacy policy and the one generated by our approach.}

\newtext{While ChatGPT demonstrates potential in generating privacy policies, its limitations arise from the lack of adequate information provided to it. Without clear guiding prompts and relevant information, ChatGPT's ability to produce a privacy policy that fully complies with basic legal requirements is constrained. This highlights the significance of our approach and why it is preferable over using ChatGPT, as our PPGen framework includes and helps cover a wide range of essential details in a privacy policy. In addition, our approach involves privacy policymakers, considers legal requirements, and takes into account the specific practices of the organisation. The identified limitation prompts us to explore the fourth setting.}

\begin{tcolorbox}[boxsep=0pt,colback=white,colframe=black!75!black]
  \textbf{In comparing PPGen with ChatGPT for generating privacy policies, it was found that while ChatGPT could produce a generic high-level privacy policy, it lacks specific details and depth, resulting in lower readability, completeness, and coverage.}
\end{tcolorbox}

\subsubsection{\newtext{RQ4. PPGen with ChatGPT}} \newtext{Following rules in our Engine, 107 questions were posed to ChatGPT for this evaluation. Out of these, 86 questions received correct responses (80.37\%), while 21 questions were answered incorrectly. The inaccuracies in responses can be categorised into two scenarios. In the first scenario, ChatGPT provided incorrect answers because the information sought was not explicitly presented in the provided privacy policy. For instance, when asked, ``Does Moodle have a controller representative?'' ChatGPT responded with, ``Yes, based on the provided privacy policy, Moodle does mention a Data Protection Officer (DPO) as a contact point for privacy-related matters...''. This response was incorrect, as Moodle did not include information about a controller representative in its privacy policy. In this case, ChatGPT was unable to find the term ``controller representative''. It then selected a similar term, Data Protection Officer, from the privacy policy as the answer. The second scenario occurred when ChatGPT treated similar actions uniformly. For instance, three closed questions pertained to sending emails to request access, deletion, and rectification of personal data. Despite Moodle not requiring users to send emails for these actions, ChatGPT misunderstood and generated identical responses for all three questions, altering only one sentence to align with each specific right. While ChatGPT shows potential for generating privacy policies and assisting privacy policymakers, limitations exist. An active involvement from privacy policymakers remains crucial to ensure accuracy and alignment with legal requirements.}

\begin{tcolorbox}[boxsep=0pt,colback=white,colframe=black!75!black]
  \textbf{ChatGPT can assist privacy policymakers in answering questions in PPGen, thereby reducing the time needed to generate a privacy policy. However, privacy policymakers are still required to check for accuracy and alignment with legal requirements.}
\end{tcolorbox}

%% file: sections/criteria1-results.tex
\begin{comment}
\begin{table}[ht]
	\centering
	\caption{The evaluation results on readability of Moodle privacy policies against existing privacy policy generators.}
	\label{tab:criteria1-results}
	\footnotesize
	\resizebox{3.25in}{!}{%
		\begin{tabular}{@{}lccccc@{}}
			\toprule
			\multicolumn{1}{c}{\multirow{2}{*}{\textbf{Metric}}}          & \multicolumn{5}{c}{\textbf{Privacy policy generators}}                                                                                                                                                        \\ \cmidrule(l){2-6} 
			\multicolumn{1}{c}{}  & \textbf{\begin{tabular}[c]{@{}c@{}}Moodle\\ (original)\end{tabular}} & \textbf{\begin{tabular}[c]{@{}c@{}}PPGen\\ (our approach)\end{tabular}} & \textbf{\begin{tabular}[c]{@{}c@{}}Privacy-\\ Policies\end{tabular}} & \textbf{Termly} & \textbf{\begin{tabular}[c]{@{}c@{}}Terms-\\ Feed\end{tabular}} \\ \midrule
			\#words  & 2,475 & 2,578  & 2,449 & 5,293  & 2,450    \\
			\#sentences  & 85  & 92   & 89   & 207   & 89     \\
			%\#words per sentence & 19.8 & 22.4 & 24.3 & 19.3 & 24.3 \\
			FRE score & 45.5    & 37.4  & 37.3 & 39  & 37.4   \\
			Reading time  & 9m 3s   & 9m 19s & 8m 57s & 19m 25s & 8m 57s   \\ \bottomrule
		\end{tabular}%
	}
\end{table}
\end{comment}

\begin{table}[ht]
    \centering
    \caption{The evaluation results on readability of \\Moodle privacy policies.}
    \label{tab:criteria1-results}
    \footnotesize
    \resizebox{3.25in}{!}{%
        \begin{tabular}{@{}lcccccc@{}}
            \toprule
            \multicolumn{1}{c}{\multirow{2}{*}{\textbf{Metric}}} & \multicolumn{6}{c}{\textbf{Privacy policy generators}}  \\ \cmidrule(l){2-7}
            \multicolumn{1}{c}{}  & 
            \multicolumn{1}{c}\textbf{\begin{tabular}[c]{@{}c@{}}Moodle\\(original)\end{tabular}} & \multicolumn{1}{c}\textbf{\begin{tabular}[c]{@{}c@{}}PPGen\\(our approach)\end{tabular}} & \multicolumn{1}{c}\textbf{\begin{tabular}[c]{@{}c@{}}Privacy-\\Policies\end{tabular}} & \multicolumn{1}{c}\textbf{Termly} & \multicolumn{1}{c}\textbf{\begin{tabular}[c]{@{}c@{}}Terms-\\Feed\end{tabular}} & \multicolumn{1}{c}\textbf{\newtext{ChatGPT}} \\ \midrule
            \#words & 2,475 & 2,578 & 2,449  & 5,293  & 2,450 & \newtext{592}  \\
            \#sentences & 85  & 92  & 89  & 207   & 89 & \newtext{26} \\
            %\#words per sentence & 19.8 & 22.4 & 24.3 & 19.3 & 24.3 & 15 \\
            FRE score & 45.5 & 37.4 & 37.3 & 39 & 37.4 & \newtext{19.3} \\
            Reading time & 9m 3s & 9m 19s & 8m 57s  & 19m 25s & 8m 57s & \newtext{2m 39s} \\ \bottomrule
        \end{tabular}%
    }
\end{table}

%% file: sections/criteria2-results.tex
\begin{table*}[ht]
    \centering
    \caption{Some examples of completeness checking and their evaluation results of Moodle privacy policy.}
    \label{tab:criteria2-results}
    
    \resizebox{7in}{!}{%
        \begin{tabular}{@{}clcccccc@{}}
            \toprule
            \multirow{2}{*}{\textbf{ID}} & \multicolumn{1}{c}{\multirow{2}{*}{\textbf{\begin{tabular}[c]{@{}c@{}}Completeness criteria\\ (pseudo-code statements)\\ {[}precondition{]}, \textless{}postcondition\textgreater{}\end{tabular}}}}                   & \multicolumn{6}{c}{\textbf{Privacy policy generators}}\\ \cmidrule(l){3-8} 
            & \multicolumn{1}{c}{} & \textbf{\begin{tabular}[c]{@{}c@{}}Moodle\\ (original)\end{tabular}} & \textbf{\begin{tabular}[c]{@{}c@{}}PPGen\\ (our approach)\end{tabular}} & \textbf{PrivacyPolicies}                                             & \textbf{Termly}                                                      & \textbf{TermsFeed} & \textbf{\newtext{ChatGPT}}                        \\ \midrule
            C2                           & \begin{tabular}[c]{@{}l@{}}{[} {]}, \textless{}CONTROLLER.CONTACT\{LEGAL ADDRESS, EMAIL, or PHONE NUMBER\} must be identified\textgreater{}\end{tabular}                                                            & Satisfied                                                            & Satisfied                                                                   & Satisfied                                                            & Satisfied                                                            & Satisfied   &  \newtext{Satisfied}                                                        \\ \midrule
            C3                           & \begin{tabular}[c]{@{}l@{}}{[}\textit{if} CONTROLLER is located outside of Europe{]}, \textless{}\textit{then} CONTROLLER REPRESENTATIVE.IDENTITY\\\{REGISTER NUMBER, or LEGAL NAME\} must be identified\textgreater{}\end{tabular} & Unsatisfied                                                          & Unsatisfied                                                                 & Unsatisfied                                                          & Unsatisfied                                                          & Unsatisfied  & \newtext{Unsatisfied}                                                        \\ \midrule
            C6                           & \begin{tabular}[c]{@{}l@{}}{[}\textit{if} DATA SUBJECT RIGHT.COMPLAINT is identified{]}, \\ \textless{}\textit{then} DATA SUBJECT RIGHT.COMPLAINT.SA should be identified\textgreater{}\end{tabular}                                    & Satisfied                                                            & Satisfied                                                                   & \begin{tabular}[c]{@{}c@{}}Precondition\\ not satisfied\end{tabular} & Unsatisfied                                                          & \begin{tabular}[c]{@{}c@{}}Precondition\\ not satisfied\end{tabular}  & \newtext{Unsatisfied}  \\ \midrule
            C15                          & \begin{tabular}[c]{@{}l@{}}{[}\textit{if} personal data is collected indirectly{]},\textless{}\textit{then} PD ORIGIN.INDIRECT must be identified\textgreater{}\end{tabular}                              & \begin{tabular}[c]{@{}c@{}}Precondition\\ not satisfied\end{tabular} & \begin{tabular}[c]{@{}c@{}}Precondition\\ not satisfied\end{tabular}        & \begin{tabular}[c]{@{}c@{}}Precondition\\ not satisfied\end{tabular} & \begin{tabular}[c]{@{}c@{}}Precondition\\ not satisfied\end{tabular} & \begin{tabular}[c]{@{}c@{}}Precondition\\ not satisfied\end{tabular} & \begin{tabular}[c]{@{}c@{}}\newtext{Satisfied but not}\\\newtext{aligned with original}\end{tabular} \\ \midrule
            C16                          & \begin{tabular}[c]{@{}l@{}}{[}\textit{if} PD ORIGIN.INDIRECT is identified{]}, \\ \textless{}\textit{then} PD ORIGIN.INDIRECT.\{THIRD PARTY, or PUBLICLY\} should be identified\textgreater{}\end{tabular}                              & \begin{tabular}[c]{@{}c@{}}Precondition\\ not satisfied\end{tabular} & \begin{tabular}[c]{@{}c@{}}Precondition\\ not satisfied\end{tabular} & \begin{tabular}[c]{@{}c@{}}Precondition\\ not satisfied\end{tabular} & \begin{tabular}[c]{@{}c@{}}Precondition\\ not satisfied\end{tabular} & \begin{tabular}[c]{@{}c@{}}Precondition\\ not satisfied\end{tabular} & \begin{tabular}[c]{@{}c@{}}\newtext{Satisfied but not}\\\newtext{aligned with original}\end{tabular} \\ \bottomrule
        \end{tabular}%
    }
\end{table*}

%% file: sections/criteria1-results-ii.tex
\begin{table}[ht]
    \centering
    \caption{The evaluation results on readability of \\Brightspace privacy policies.}
    \label{tab:criteria1-results-ii}
    \footnotesize
        \begin{tabular}{@{}lcc@{}}
            \toprule
            \multicolumn{1}{c}{\multirow{2}{*}{\textbf{Metric}}} & \multicolumn{2}{c}{\textbf{Privacy policy generators}}  \\ \cmidrule(l){2-3}
            \multicolumn{1}{c}{}  & 
            \multicolumn{1}{c}\textbf{\begin{tabular}[c]{@{}c@{}}Brightspace\\(original)\end{tabular}} & \multicolumn{1}{c}\textbf{\begin{tabular}[c]{@{}c@{}}PPGen\\(our approach)\end{tabular}} \\ \midrule
            \#words & 3,841 & 3,038 \\
            \#sentences & 129  & 108   \\
            FRE score & 28.9 & 39.3  \\
            Reading time & 18m 42s & 15m 25s  \\ \bottomrule
        \end{tabular}
\end{table}

%% file: sections/criteria2-results-ii.tex
\begin{table*}[ht]
    \centering
    \caption{Some examples of completeness checking and their evaluation results of Brightspace privacy policy.}
    \label{tab:criteria2-results-ii}
    
    \resizebox{5.5in}{!}{%
        \begin{tabular}{@{}clcc@{}}
            \toprule
            \multirow{2}{*}{\textbf{ID}} & \multicolumn{1}{c}{\multirow{2}{*}{\textbf{\begin{tabular}[c]{@{}c@{}}Completeness criteria\\ (pseudo-code statements)\\ {[}precondition{]}, \textless{}postcondition\textgreater{}\end{tabular}}}}                   & \multicolumn{2}{c}{\textbf{Privacy policy generators}} \\ \cmidrule(l){3-4} 
            & \multicolumn{1}{c}{}                                                           & \textbf{\begin{tabular}[c]{@{}c@{}}Brightspace\\ (original)\end{tabular}} & \textbf{\begin{tabular}[c]{@{}c@{}}PPGen\\ (our approach)\end{tabular}}  \\ \midrule
            C2                           & \begin{tabular}[c]{@{}l@{}}{[} {]}, \textless{}CONTROLLER.CONTACT\{LEGAL ADDRESS, EMAIL, or PHONE NUMBER\} must be identified\textgreater{}\end{tabular}   & Satisfied                                                            & Satisfied                                                  \\ \midrule
            C3                           & \begin{tabular}[c]{@{}l@{}}{[}\textit{if} CONTROLLER is located outside of Europe{]}, \textless{}\textit{then} CONTROLLER REPRESENTATIVE.IDENTITY\\\{REGISTER NUMBER, or LEGAL NAME\} must be identified\textgreater{}\end{tabular} & Satisfied                                                          & Satisfied      \\ \midrule
            C6                           & \begin{tabular}[c]{@{}l@{}}{[}\textit{if} DATA SUBJECT RIGHT.COMPLAINT is identified{]}, \\ \textless{}\textit{then} DATA SUBJECT RIGHT.COMPLAINT.SA should be identified\textgreater{}\end{tabular}                                    & Unsatisfied                                                            & Unsatisfied   \\ \midrule
            C15                          & \begin{tabular}[c]{@{}l@{}}{[}\textit{if} personal data is collected indirectly{]},\textless{}\textit{then} PD ORIGIN.INDIRECT must be identified\textgreater{}\end{tabular}              & Satisfied                                                          & Satisfied     \\ \midrule
            C16                          & \begin{tabular}[c]{@{}l@{}}{[}\textit{if} PD ORIGIN.INDIRECT is identified{]}, \\ \textless{}\textit{then} PD ORIGIN.INDIRECT.\{THIRD PARTY, or PUBLICLY\} should be identified\textgreater{}\end{tabular}  & Satisfied                                                          & Satisfied  \\ \bottomrule
        \end{tabular}%
    }
\end{table*}

%% file: sections/7-threats.tex
\section{\newtext{Threats to validity}} \label{sec:threats}

\subsection{Internal}

\newtext{The first author performed all the steps in privacy clause development, privacy policy generation and evaluation, with some subjective judgements. To mitigate these threats, all the steps were transparently documented, reviewed by the other authors and replicable. The author used four privacy policy generators (i.e., PPGen, PrivacyPolicies, Termly and Termsfeed) to generate the privacy policies based on Moodle. Evaluation results confirmed alignment between the original and PPGen generated privacy policies. It also confirmed that our approach outperformed other privacy policy generators on completeness and coverage. In addition, the evaluation was conducted using the unseen OLA, Brightspace, whose privacy policy was not used in the privacy clause development process. This evaluation confirmed that our approach produced a privacy policy aligned with Brightspace's original policy. We acknowledge that our methodology's reliance on developer or privacy policymaker input. This can lead to some errors or inconsistencies if questions are misunderstood or inaccurately answered.}

\subsection{External}

\newtext{We used ten privacy policies of OLAs as input. We acknowledge that this set of privacy policies may not be representative of other privacy policies. However, the privacy clauses extracted from those privacy policies are sufficient to identify both common and unique clauses for metadata types. Common clauses were refined into one clause, while the unique clauses were retained in the privacy clause refinement. In addition, privacy clause extraction required significant effort (i.e., 8-10 person-hours depending on the length of privacy policies). Further investigation is required to explore privacy policies in different domains. In addition, it is important to note that our research relies on existing privacy clauses from selected OLAs. However, this reliance may not fully capture a wide range of privacy concerns presented in diverse online learning environments.}

Our study was conducted without input from legal experts. However, the conceptual model of privacy-policy metadata and completeness checking criteria we adopted \cite{Amaral2021} were reviewed, validated and evaluated by legal experts. The privacy policy generated by our PPGen is not intended to be a substitute for legal advice. Organisations may require further legal or professional advice on any special requirements or specific questions.

\newtext{Our study primarily focuses on GDPR compliance owing to its worldwide coverage, significance, and profound impact on organisations. GDPR stands as a cornerstone in the field of data protection regulations, setting rigorous standards for the handling and processing of personal data. Many countries have developed or adopted data privacy laws based on or comparable to GDPR \cite{Simmons2022}. Focusing on a particular regulation is crucial for gaining deeper insight into its functionality and efficacy. However, we acknowledge the potential limitations in applicability that may arise in jurisdictions with different or additional data protection requirements. Our future work will focus on ensuring the compatibility of our system with various data protection requirements.}

\newtext{We acknowledge that the paper does not discuss the system's potential to adapt to changes in GDPR or other data protection laws. This limitation could impact the long-term viability of the generated policies and introduce potential constraints in scalability and flexibility as privacy laws evolve over time. Hence, updates to the rule-based system are necessary to ensure ongoing compliance. In future work, we intend to address this limitation by leveraging artificial intelligence technologies to regularly update or expand the system's library, thereby reflecting the evolving landscape of privacy laws and regulatory changes.}

%% file: sections/8-conclusion.tex
\section{Conclusions and future work} \label{sec:conclusion}

We proposed a new approach to support automated generation of privacy policies for software applications. \newtext{We applied and demonstrated our approach using OLAs.} We extracted 1,081 privacy clauses from 10 privacy policies based on the privacy-policy metadata types \cite{Amaral2021}. The extracted clauses were used to build a library of privacy clauses.
We developed PPGen, an interactive system that helps software developers in generating a customised privacy policy. It uses answers to 167 questions to determine the privacy clauses that are included in the privacy policy generated by the system. A privacy policy template is used to indicate where the privacy clauses need to be placed. In addition, we added 58 non-compliant and 14 warning clauses to inform OLA developers of non-compliances and points that need to be reviewed in the generated privacy policy respectively.  
\newtext{We conducted evaluations in four different settings. In the first setting,} we compared PPGen generated privacy policy to four others on readability, completeness and coverage. PPGen generated privacy policy is the most complete and comprehensive compared to privacy policies generated by other policy generators, however, its readability score ranks third among those five privacy policies. \newtext{In the second setting, we employed our approach to generating a privacy policy for a new OLA that was not included in the clause extraction process. The results confirmed our approach's capability to generate privacy policies for a new software application. In the third setting, we evaluated the privacy policy generated by our approach against the privacy policy generated by ChatGPT. The results showed that our approach outperformed ChatGPT in all criteria for privacy policy generation. In the final setting, we examined the use of ChatGPT with our approach. The results showed the potential of using ChatGPT to assist privacy policymakers. However, the involvement of privacy policymakers remains crucial for privacy policy generation.}

\newtext{Our future work involves exploring the use of Generative Artificial Intelligence to enable our Engine component to automatically update and expand the library of privacy clauses. In addition, we aim to develop the Generator component to answer the questions without requiring inputs from privacy policymakers.}

%Our future work involves exploring the use of Generative Artificial Intelligence to enable our Engine component to automatically answer the questions without getting responses from privacy policymakers. We plan to adapt and fine tune a question-answering system (e.g., ChatGPT \cite{OpenAI}) to automate this process.

%Our future work involves building a fully automated QA system that uses the knowledge learnt from issue reports to answer the questions without using any inputs from a software developer or a privacy policymaker. We plan to adopt an artificial intelligence chatbot to support this automation. This tool can be integrated into existing issue tracking systems such as JIRA\footnote{https://www.atlassian.com/software/jira} to help generate a customised and GDPR-compliant privacy policy. In addition, we plan to improve the generalisability and extend the evaluation of our approach.
%We plan to cover other privacy regulations and applicable to software applications in other domains apart from the OLAs.

%% file: policy-generation.bbl
% Generated by IEEEtranN.bst, version: 1.13 (2008/09/30)
\begin{thebibliography}{53}
\providecommand{\natexlab}[1]{#1}
\providecommand{\url}[1]{#1}
\csname url@samestyle\endcsname
\providecommand{\newblock}{\relax}
\providecommand{\bibinfo}[2]{#2}
\providecommand{\BIBentrySTDinterwordspacing}{\spaceskip=0pt\relax}
\providecommand{\BIBentryALTinterwordstretchfactor}{4}
\providecommand{\BIBentryALTinterwordspacing}{\spaceskip=\fontdimen2\font plus
\BIBentryALTinterwordstretchfactor\fontdimen3\font minus
  \fontdimen4\font\relax}
\providecommand{\BIBforeignlanguage}[2]{{%
\expandafter\ifx\csname l@#1\endcsname\relax
\typeout{** WARNING: IEEEtranN.bst: No hyphenation pattern has been}%
\typeout{** loaded for the language `#1'. Using the pattern for}%
\typeout{** the default language instead.}%
\else
\language=\csname l@#1\endcsname
\fi
#2}}
\providecommand{\BIBdecl}{\relax}
\BIBdecl

\bibitem[{Office Journal of the European
  Union}(2016)]{OfficeJournaloftheEuropeanUnion;2016}
\BIBentryALTinterwordspacing
{Office Journal of the European Union}, ``{General Data Protection Regulation
  (GDPR)},'' Tech. Rep., 2016. [Online]. Available:
  \url{https://eur-lex.europa.eu/legal-content/EN/TXT/PDF/?uri=CELEX:32016R0679}
\BIBentrySTDinterwordspacing

\bibitem[{State of California Department of
  Justice}(2018)]{StateofCaliforniaDepartmentofJustice2018}
\BIBentryALTinterwordspacing
{State of California Department of Justice}, ``{California Consumer Privacy Act
  (CCPA)},'' 2018. [Online]. Available: \url{https://oag.ca.gov/privacy/ccpa}
\BIBentrySTDinterwordspacing

\bibitem[Amaral et~al.(2021{\natexlab{a}})Amaral, Abualhaija, Torre,
  Sabetzadeh, and Briand]{Amaral2021}
O.~Amaral, S.~Abualhaija, D.~Torre, M.~Sabetzadeh, and L.~Briand, ``{AI-enabled
  Automation for Completeness Checking of Privacy Policies},'' \emph{IEEE
  Transactions on Software Engineering}, vol.~48, no.~11, pp. 4647--4674, 2021.

\bibitem[Tahaei et~al.(2023)Tahaei, Vaniea, and Rashid]{Tahaei2023}
\BIBentryALTinterwordspacing
M.~Tahaei, K.~Vaniea, and A.~Rashid, ``{Embedding Privacy Into Design Through
  Software Developers: Challenges and Solutions},'' \emph{IEEE Security and
  Privacy}, vol.~21, no.~1, pp. 49--57, 2023. [Online]. Available:
  \url{www.computer.org/security}
\BIBentrySTDinterwordspacing

\bibitem[Guntamukkala et~al.(2016)Guntamukkala, Dara, and
  Grewal]{Guntamukkala2016}
N.~Guntamukkala, R.~Dara, and G.~Grewal, ``{A machine-learning based approach
  for measuring the completeness of online privacy policies},'' in
  \emph{Proceedings - 2015 IEEE 14th International Conference on Machine
  Learning and Applications, ICMLA 2015}, 2016, pp. 289--294.

\bibitem[Tesfay et~al.(2018)Tesfay, Hofmann, Nakamura, Kiyomoto, and
  Serna]{Tesfay2018}
W.~B. Tesfay, P.~Hofmann, T.~Nakamura, S.~Kiyomoto, and J.~Serna,
  ``{Privacyguide: Towards an implementation of the EU GDPR on internet privacy
  policy evaluation},'' in \emph{IWSPA 2018 - Proceedings of the 4th ACM
  International Workshop on Security and Privacy Analytics, Co-located with
  CODASPY 2018}.\hskip 1em plus 0.5em minus 0.4em\relax ACM, 2018, pp. 15--21.

\bibitem[Torre et~al.(2020)Torre, Abualhaija, Sabetzadeh, Briand, Baetens,
  Goes, and Forastier]{Torre}
D.~Torre, S.~Abualhaija, M.~Sabetzadeh, L.~Briand, K.~Baetens, P.~Goes, and
  S.~Forastier, ``{An AI-assisted Approach for Checking the Completeness of
  Privacy Policies Against GDPR},'' \emph{2020 IEEE 28th International
  Requirements Engineering Conference (RE)}, pp. 136--146, 2020.

\bibitem[M{\"{u}}ller et~al.(2019)M{\"{u}}ller, Kowatsch, Debus, Mirdita, and
  B{\"{o}}ttinger]{Muller2019}
N.~M. M{\"{u}}ller, D.~Kowatsch, P.~Debus, D.~Mirdita, and K.~B{\"{o}}ttinger,
  ``{On GDPR Compliance of Companies' Privacy Policies},'' in \emph{Lecture
  Notes in Computer Science (including subseries Lecture Notes in Artificial
  Intelligence and Lecture Notes in Bioinformatics)}, sep 2019, vol. 11697, pp.
  151--159.

\bibitem[Anthonysamy et~al.(2014)Anthonysamy, Greenwood, and
  Rashid]{Anthonysamy2014}
P.~Anthonysamy, P.~Greenwood, and A.~Rashid, ``{A method for analysing
  traceability between privacy policies and privacy controls of online social
  networks},'' \emph{Lecture Notes in Computer Science (including subseries
  Lecture Notes in Artificial Intelligence and Lecture Notes in
  Bioinformatics)}, vol. 8319, pp. 187--202, 2014.

\bibitem[Termly(2022)]{Termly2022}
\BIBentryALTinterwordspacing
Termly, ``{Free Privacy Policy Generator - Create a Privacy Policy},'' 2022.
  [Online]. Available:
  \url{https://termly.io/products/privacy-policy-generator/}
\BIBentrySTDinterwordspacing

\bibitem[PrivacyPolicies(2022)]{PrivacyPolicies2022}
\BIBentryALTinterwordspacing
PrivacyPolicies, ``{Welcome to PrivacyPolicies.com - PrivacyPolicies},'' 2022.
  [Online]. Available: \url{https://app.privacypolicies.com/}
\BIBentrySTDinterwordspacing

\bibitem[TermsFeed(2023)]{TermsFeed2023}
\BIBentryALTinterwordspacing
TermsFeed, ``{Free Privacy Policy Generator - TermsFeed},'' 2023. [Online].
  Available: \url{https://www.termsfeed.com/privacy-policy-generator/}
\BIBentrySTDinterwordspacing

\bibitem[Sangaroonsilp et~al.(2024)Sangaroonsilp, Dam, Haggag, and
  Grundy]{reppkg-privacygen}
\BIBentryALTinterwordspacing
P.~Sangaroonsilp, H.~K. Dam, O.~Haggag, and J.~Grundy, ``{Replication package
  for ''Interactive GDPR-Compliant Privacy Policy Generation for Software
  Applications'' manuscript},'' 2024. [Online]. Available:
  \url{https://drive.google.com/drive/folders/1ltUTPGj5GBR_iOW5qYU_fnohDY2MdkQi}
\BIBentrySTDinterwordspacing

\bibitem[Sun and Xue(2020)]{Sun2020}
R.~Sun and M.~Xue, ``{Quality Assessment of Online Automated Privacy Policy
  Generators: An Empirical Study},'' in \emph{ACM International Conference
  Proceeding Series}, 2020, pp. 270--275.

\bibitem[Yu et~al.(2017)Yu, Zhang, Luo, Xue, and Chang]{Yu2017}
L.~Yu, T.~Zhang, X.~Luo, L.~Xue, and H.~Chang, ``{Toward Automatically
  Generating Privacy Policy for Android Apps},'' \emph{IEEE Transactions on
  Information Forensics and Security}, vol.~12, no.~4, pp. 865--880, apr 2017.

\bibitem[Yu et~al.(2015)Yu, Zhang, Luo, and Xue]{Yu2015a}
L.~Yu, T.~Zhang, X.~Luo, and L.~Xue, ``{AutoPPG: Towards automatic generation
  of privacy policy for android applications},'' in \emph{SPSM 2015 -
  Proceedings of the 5th Annual ACM CCS Workshop on Security and Privacy in
  Smartphones and Mobile Devices, co-located with: CCS 2015}, 2015, pp. 39--50.

\bibitem[Yu et~al.(2021)Yu, Luo, Chen, Zhou, Zhang, Chang, and Leung]{Yu2021}
L.~Yu, X.~Luo, J.~Chen, H.~Zhou, T.~Zhang, H.~Chang, and H.~K. Leung,
  ``{PPChecker: Towards Accessing the Trustworthiness of Android Apps' Privacy
  Policies},'' \emph{IEEE Transactions on Software Engineering}, vol.~47,
  no.~2, pp. 221--242, feb 2021.

\bibitem[Miao(2014)]{Miao2014}
D.~Y. Miao, ``{PrivacyInformer : An Automated Privacy Description Generator for
  the MIT App Inventor},'' Ph.D. dissertation, Massachusetts Institute of
  Technology, 2014.

\bibitem[Rowan and Dehlinger(2014)]{Rowan2014}
\BIBentryALTinterwordspacing
M.~Rowan and J.~Dehlinger, ``{Encouraging privacy by design concepts with
  privacy policy auto-generation in eclipse (PAGE)},'' in \emph{ETX 2014 -
  Proceedings of the 2014 ACM SIGPLAN Workshop on Eclipse Technology eXchange,
  Part of SPLASH 2014}.\hskip 1em plus 0.5em minus 0.4em\relax Association for
  Computing Machinery, Inc, oct 2014, pp. 9--14. [Online]. Available:
  \url{http://dx.doi.org/10.1145/2688130.2688134}
\BIBentrySTDinterwordspacing

\bibitem[Shopify(2022)]{Shop}
\BIBentryALTinterwordspacing
Shopify, ``{Privacy Policy Generator | GDPR Privacy Policy Generator {\&} Free
  Template},'' 2022. [Online]. Available:
  \url{https://www.shopify.com/tools/policy-generator}
\BIBentrySTDinterwordspacing

\bibitem[Brilliant(2020)]{Brilliant2020}
\BIBentryALTinterwordspacing
Brilliant, ``{Brilliant.org Privacy Statement},'' 2020. [Online]. Available:
  \url{https://brilliant.org/privacy/}
\BIBentrySTDinterwordspacing

\bibitem[Coursera(2022)]{Coursera}
\BIBentryALTinterwordspacing
Coursera, ``{Privacy Notice | Coursera},'' 2022. [Online]. Available:
  \url{https://www.coursera.org/about/privacy}
\BIBentrySTDinterwordspacing

\bibitem[EdX(2021)]{edX2021}
\BIBentryALTinterwordspacing
EdX, ``{Privacy Policy},'' 2021. [Online]. Available:
  \url{https://www.edx.org/edx-privacy-policy}
\BIBentrySTDinterwordspacing

\bibitem[FutureLearn(2022)]{FutureLearn2022}
\BIBentryALTinterwordspacing
FutureLearn, ``{Privacy policy},'' 2022. [Online]. Available:
  \url{https://www.futurelearn.com/info/terms/privacy-policy}
\BIBentrySTDinterwordspacing

\bibitem[MasterClass(2020)]{MasterClass2020}
\BIBentryALTinterwordspacing
MasterClass, ``{Privacy Policy},'' 2020. [Online]. Available:
  \url{https://privacy.masterclass.com/policies}
\BIBentrySTDinterwordspacing

\bibitem[Mindvalley(2021)]{Mindvalley2021}
\BIBentryALTinterwordspacing
Mindvalley, ``{Privacy Policy},'' 2021. [Online]. Available:
  \url{https://www.mindvalley.com/privacy-policy}
\BIBentrySTDinterwordspacing

\bibitem[Moodle(2022)]{Moodlea}
\BIBentryALTinterwordspacing
Moodle, ``{Privacy Notice | Moodle},'' 2022. [Online]. Available:
  \url{https://moodle.com/privacy-notice/}
\BIBentrySTDinterwordspacing

\bibitem[Skillshare(2019)]{Skillshare2019}
\BIBentryALTinterwordspacing
Skillshare, ``{Privacy Policy},'' 2019. [Online]. Available:
  \url{https://www.skillshare.com/en/privacy}
\BIBentrySTDinterwordspacing

\bibitem[Udacity(2022)]{Udacity2022}
\BIBentryALTinterwordspacing
Udacity, ``{Udacity Privacy Policy},'' 2022. [Online]. Available:
  \url{https://www.udacity.com/legal/en-us/privacy}
\BIBentrySTDinterwordspacing

\bibitem[Udemy(2021)]{Udemy2021}
\BIBentryALTinterwordspacing
Udemy, ``{Privacy Policy},'' 2021. [Online]. Available:
  \url{https://www.udemy.com/terms/privacy/}
\BIBentrySTDinterwordspacing

\bibitem[Amaral et~al.(2021{\natexlab{b}})Amaral, Abualhaija, Torre,
  Sabetzadeh, and Briand]{Amaral}
\BIBentryALTinterwordspacing
O.~Amaral, S.~Abualhaija, D.~Torre, M.~Sabetzadeh, and L.~Briand, ``{Glossary
  and Completeness Criteria traceability to the GDPR articles},'' 2021.
  [Online]. Available:
  \url{https://dropit.uni.lu/invitations/?share=922b2cf938f4304861b8}
\BIBentrySTDinterwordspacing

\bibitem[Digital(2023)]{Digital}
\BIBentryALTinterwordspacing
Digital, ``{The Best Privacy Policy Generators of 2023},'' 2023. [Online].
  Available: \url{https://digital.com/best-privacy-policy-generators/}
\BIBentrySTDinterwordspacing

\bibitem[{State of California Department of Justice}(2020)]{CPRA2020}
\BIBentryALTinterwordspacing
{State of California Department of Justice}, ``{California Consumer Privacy Act
  of 2018},'' 2020. [Online]. Available:
  \url{https://leginfo.legislature.ca.gov/faces/codes_displayText.xhtml?division=3.&part=4.&lawCode=CIV&title=1.81.5}
\BIBentrySTDinterwordspacing

\bibitem[{Federal Trade Commision}(2000)]{COPPA2000}
\BIBentryALTinterwordspacing
{Federal Trade Commision}, ``{Children's Online Privacy Protection Rule
  ("COPPA")},'' 2000. [Online]. Available:
  \url{https://www.ftc.gov/legal-library/browse/rules/childrens-online-privacy-protection-rule-coppa}
\BIBentrySTDinterwordspacing

\bibitem[{Privacy Commisioner of Canada}(2019)]{PIPEDA2019}
\BIBentryALTinterwordspacing
{Privacy Commisioner of Canada}, ``{Personal Information Protection and
  Electronic Documents Act},'' Tech. Rep., 2019. [Online]. Available:
  \url{https://laws-lois.justice.gc.ca/PDF/P-8.6.pdf}
\BIBentrySTDinterwordspacing

\bibitem[{European Commission}(2017)]{ePrivacy2017}
\BIBentryALTinterwordspacing
{European Commission}, ``{Proposal for an ePrivacy Regulation},'' 2017.
  [Online]. Available:
  \url{https://digital-strategy.ec.europa.eu/en/policies/eprivacy-regulation}
\BIBentrySTDinterwordspacing

\bibitem[D2L(2024)]{D2L2024}
\BIBentryALTinterwordspacing
D2L, ``{Brightspace Learning Management System},'' 2024. [Online]. Available:
  \url{https://www.d2l.com/en-apac/brightspace/}
\BIBentrySTDinterwordspacing

\bibitem[D2L(2023)]{Brightspace}
\BIBentryALTinterwordspacing
------, ``{Privacy Policy},'' 2023. [Online]. Available:
  \url{https://www.d2l.com/en-apac/legal/privacy/}
\BIBentrySTDinterwordspacing

\bibitem[OpenAI(2023{\natexlab{a}})]{OpenAI}
\BIBentryALTinterwordspacing
OpenAI, ``{OpenAI},'' 2023. [Online]. Available: \url{https://openai.com/}
\BIBentrySTDinterwordspacing

\bibitem[{Krystal Hu}(2023)]{KrystalHu2023}
\BIBentryALTinterwordspacing
{Krystal Hu}, ``{ChatGPT sets record for fastest-growing user base - analyst
  note},'' 2023. [Online]. Available:
  \url{https://www.reuters.com/technology/chatgpt-sets-record-fastest-growing-user-base-analyst-note-2023-02-01/}
\BIBentrySTDinterwordspacing

\bibitem[Marr(2023)]{Marr2023}
\BIBentryALTinterwordspacing
B.~Marr, ``{A short history of ChatGPT: How we got to where we are today},''
  2023. [Online]. Available:
  \url{https://www.forbes.com/sites/bernardmarr/2023/05/19/a-short-history-of-chatgpt-how-we-got-to-where-we-are-today/?sh=5d096667674f}
\BIBentrySTDinterwordspacing

\bibitem[OpenAI(2023{\natexlab{b}})]{ChatGPT}
\BIBentryALTinterwordspacing
OpenAI, ``{ChatGPT},'' 2023. [Online]. Available:
  \url{https://openai.com/chatgpt}
\BIBentrySTDinterwordspacing

\bibitem[Noy and Zhang(2023)]{Noy2023}
\BIBentryALTinterwordspacing
S.~Noy and W.~Zhang, ``{Experimental evidence on the productivity effects of
  generative artificial intelligence},'' \emph{Science}, vol. 381, no. 6654,
  pp. 187--192, jul 2023. [Online]. Available: \url{https://www.science.org}
\BIBentrySTDinterwordspacing

\bibitem[Graber et~al.(2002)Graber, D'Alessandro, and Johnson-West]{Graber2002}
M.~A. Graber, D.~M. D'Alessandro, and J.~Johnson-West, ``{Reading level of
  privacy policies on Internet health Web sites},'' \emph{Journal of Family
  Practice}, vol.~51, no.~7, pp. 642--645, 2002.

\bibitem[McDonald et~al.(2009)McDonald, Reeder, Kelley, and
  Cranor]{McDonald2009}
\BIBentryALTinterwordspacing
A.~M. McDonald, R.~W. Reeder, P.~G. Kelley, and L.~F. Cranor, ``{A comparative
  study of online privacy policies and formats},'' in \emph{Lecture Notes in
  Computer Science (including subseries Lecture Notes in Artificial
  Intelligence and Lecture Notes in Bioinformatics)}, vol. 5672 LNCS.\hskip 1em
  plus 0.5em minus 0.4em\relax Springer, Berlin, Heidelberg, 2009, pp. 37--55.
  [Online]. Available:
  \url{https://link.springer.com/chapter/10.1007/978-3-642-03168-7_3}
\BIBentrySTDinterwordspacing

\bibitem[Savla and Martino(2012)]{Savla2012}
P.~Savla and L.~D. Martino, ``{Content analysis of privacy policies for health
  social networks},'' in \emph{Proceedings - 2012 IEEE International Symposium
  on Policies for Distributed Systems and Networks, POLICY 2012}, 2012, pp.
  94--101.

\bibitem[Krumay and Klar(2020)]{Krumay2020}
\BIBentryALTinterwordspacing
B.~Krumay and J.~Klar, ``{Readability of privacy policies},'' in \emph{Lecture
  Notes in Computer Science (including subseries Lecture Notes in Artificial
  Intelligence and Lecture Notes in Bioinformatics)}, vol. 12122 LNCS.\hskip
  1em plus 0.5em minus 0.4em\relax Springer, 2020, pp. 388--399. [Online].
  Available:
  \url{https://link.springer.com/chapter/10.1007/978-3-030-49669-2_22}
\BIBentrySTDinterwordspacing

\bibitem[Flesch(1948)]{Flesch1948}
R.~Flesch, ``{A new readability yardstick},'' \emph{Journal of Applied
  Psychology}, vol.~32, no.~3, pp. 221--233, jun 1948.

\bibitem[Microsoft(2022)]{Microsoft2022}
\BIBentryALTinterwordspacing
Microsoft, ``{Get your document's readability and level statistics},'' 2022.
  [Online]. Available:
  \url{https://support.microsoft.com/en-us/office/get-your-document-s-readability-and-level-statistics-85b4969e-e80a-4777-8dd3-f7fc3c8b3fd2#__toc342546557}
\BIBentrySTDinterwordspacing

\bibitem[{International Commissioner's Office}(2023)]{ICO2023}
\BIBentryALTinterwordspacing
{International Commissioner's Office}, ``{Make your own privacy notice},''
  2023. [Online]. Available:
  \url{https://ico.org.uk/for-organisations/sme-web-hub/make-your-own-privacy-notice/}
\BIBentrySTDinterwordspacing

\bibitem[Wolford(2019)]{Wolford2019}
\BIBentryALTinterwordspacing
B.~Wolford, ``{Writing a GDPR-compliant privacy notice},'' 2019. [Online].
  Available: \url{https://gdpr.eu/privacy-notice/}
\BIBentrySTDinterwordspacing

\bibitem[{PwC Australia}(2023)]{PwCAustralia2023}
\BIBentryALTinterwordspacing
{PwC Australia}, ``{Privacy Policy Template},'' 2023. [Online]. Available:
  \url{https://www.pwc.com.au/about-us/social-impact/privacy-guidelines-for-not-for-profits-nfp.html}
\BIBentrySTDinterwordspacing

\bibitem[Simmons(2022)]{Simmons2022}
\BIBentryALTinterwordspacing
D.~Simmons, ``{17 Countries with GDPR-like Data Privacy Laws},'' 2022.
  [Online]. Available:
  \url{https://insights.comforte.com/countries-with-gdpr-like-data-privacy-laws}
\BIBentrySTDinterwordspacing

\end{thebibliography}
